\documentclass[twoside]{JHEP3}
\title{Absorption Lengths in the Holographic Plasma}
\author{Irene Amado,${}^a$ Carlos Hoyos,${}^b$ Karl Landsteiner${}^a$ and Sergio
Montero${}^a$\\
  ${}^a$Instituto de F\'{\i}sica Te\'orica, C-XVI Universidad Aut\'onoma de
Madrid\\
  ~\,E-28049 Madrid, Spain\\
  ~\,E-mail: \email{Irene.Amado, Karl.Landsteiner, Sergio.Montero@uam.es}\\
  ${}^b$Department of Physics,
  Swansea University\\
  ~\,Swansea, SA2 8PP, UK\\
  ~\,E-mail: \email{C.H.Badajoz@swansea.ac.uk}} \abstract{We consider
  the effect of a periodic perturbation with frequency $\omega$ on the
  holographic $\Nfour$ plasma represented by the planar AdS black
  hole. The response of the system is given by exponentially decaying
  waves. The corresponding complex wave numbers can be found by
  solving wave equations in the AdS black hole background with
  infalling boundary conditions on the horizon in an analogous way as
  in the calculation of quasinormal modes. The complex momentum
  eigenvalues have an interpretation as poles of the retarded Green's
  functions, where the inverse of the imaginary part gives an
  absorption length $\lambda$. At zero frequency we obtain the
  screening length for a static field. These are directly related to
  the glueball masses in the dimensionally reduced theory. We also
  point out that the longest screening length corresponds to an
  operator with non-vanishing \mbox{R-charge} and thus does not have
  an interpretation as a QCD$_3$ glueball.}
\preprint{IFT-UAM/CSIC-07-33} 
\usepackage[english]{babel}
\usepackage{amssymb}
\usepackage{amsfonts}
\usepackage{amsmath}
\usepackage{graphicx}
\usepackage{epsfig}
\usepackage{psfrag}
\usepackage{cite}
\def\erf#1{(\ref{#1})} 
\newcommand{\cO}{{\cal O}}

  \newcommand{\bbN}{{\mathbb N}}

  \newcommand{\bbZ}{{\mathbb Z}}

\newcommand{\bA}{{\mathbf A}}
  
\newcommand{\bE}{{\mathbf E}}

  \newcommand{\by}{{\mathbf y}}
\newcommand{\bk}{{\mathbf k}}  \newcommand{\bq}{{\mathbf q}}
\newcommand{\bx}{{\mathbf x}}

\def\bsxi{\boldsymbol{\xi}}

\newcommand{\be}{\begin{equation}} \newcommand{\ee}{\end{equation}}
\newcommand{\bea}{\begin{eqnarray}} \newcommand{\eea}{\end{eqnarray}}
\newcommand{\beann}{\begin{eqnarray*}}
  \newcommand{\eeann}{\end{eqnarray*}}
\newcommand{\bfig}{\begin{figure}} \newcommand{\efig}{\end{figure}}
\newcommand{\ba}{\begin{array}} \newcommand{\ea}{\end{array}}
\newcommand{\bcen}{\begin{center}} \newcommand{\ecen}{\end{center}}
\newcommand{\btab}{\begin{tabular}} \newcommand{\etab}{\end{tabular}}
\newcommand{\nn}{\nonumber}
\def\tr{\operatorname{tr\:}}

\def\Res{\operatorname{Res}}     \def\sign{\operatorname{sign}}
\renewcommand{\Re}{\mathop{\rm Re}}   \renewcommand{\Im}{\mathop{\rm Im}}

\newcommand{\vev}[1]{\left\langle{#1}\right\rangle}

\newcommand{\dd}{{\rm d}}
\newcommand{\e}{{\rm e}}
\newcommand{\adsfive}{$\mathop{\rm AdS_5\times{}S^5}$\,}
  \def\Nfour{{\cal
    N}\!=\!4}

\newtheorem{Proposition}{Proposition}[section]

\newtheorem{Theorem}{Theorem}[section]
\newtheorem{Lemma}{Lemma}[section]
\newtheorem{Corrolary}{Corrolary}[section]

\newcommand{\bp}{\begin{Proposition}}	\newcommand{\ep}{\end{Proposition}}
\newcommand{\bt}{\begin{Theorem}}	\newcommand{\et}{\end{Theorem}}
\newcommand{\bl}{\begin{Lemma}}		\newcommand{\el}{\end{Lemma}}
\newcommand{\bc}{\begin{Corrolary}}	\newcommand{\ec}{\end{Corrolary}}
\begin{document}

\section{Introduction} \label{sec:intro}
The AdS/CFT correspondence is a concrete realization of the the idea
that the large-$N$ limit of non-Abelian gauge theories can be
described by a dual string theory \cite{firstadscft}. In the large 't
Hooft coupling regime the dual theory admits a description in terms of
gravitational fields over a weakly curved background. More precisely
the AdS/CFT correspondence proposes an exact duality between $\Nfour$
super Yang-Mills in four dimensions and type IIB superstrings in
\adsfive. This theory is conformal, but at finite temperature
conformal symmetry is broken and the theory is in a deconfining (or
plasma) phase. The dual description corresponds to a black hole
geometry with flat horizon \cite{Witten:1998zw}.

Heavy-ion collisions at RHIC \cite{Muller:2006ee} and lattice
simulations \cite{Karsch:2001cy} indicate that QCD actually stays
strongly coupled above the deconfinement transition up to temperatures
$T\sim 2 T_c$. Therefore, it is of great interest to develop
non-perturbative tools that can describe the properties of the
strongly coupled plasma. Lattice simulations are good to describe
thermodynamical properties, but out-of-equilibrium processes are much
harder to analyze. In this context, the AdS/CFT correspondence could
provide a better framework to derive some plasma properties using
analytic methods. Although $\Nfour$ is far from QCD, some qualitative
properties of the plasma seem to be quite similar, and AdS/CFT
computations of the shear viscosity \cite{viscosity}, the energy loss
rate of a heavy quark \cite{dragforce} or the jet quenching parameter
using light-like Wilson loops \cite{jetquenching} show good agreement
with experimental data.

Many of the properties of the plasma can be studied using linear
response theory. In this approximation, small perturbations that do
not change significantly the state of the plasma are introduced. The
system then tries to restore thermal equilibrium. That involves {\em
  dissipation} if the perturbations are localized in time or {\em
  absorption} if they are localized in space.\footnote{There can also
  be diffusion effects if conserved charges are involved.} We would
like to address the latter in this work. The absorption is directly
related to spatial correlations in the equilibrium state. At high
temperatures the system is in a very disordered phase, so measures
made in different parts of the plasma give uncorrelated results. For
the same reason, a small perturbation cannot travel too far in the
plasma before being washed out by thermal fluctuations. How far this
can be depends on the details of the plasma, but in general we expect
that the characteristic absorption lengths decrease as the temperature
increases.

In the gravity dual picture the absorptive properties of the plasma
rely on the presence of a horizon. Small classical perturbations end
up falling into the horizon, either after a finite time or after
travelling a finite distance.\footnote{In AdS space the curvature acts
  effectively as a box, so they cannot escape to infinity.} The first
is described by complex values for the eigenfrequency, the quasinormal
modes, and the second by complex momentum values. Both are intimately
related, they correspond to solutions of the linearized equations of
motion. They also satisfy the same boundary conditions, Dirichlet at
the AdS boundary\,\footnote{Actually, the condition is that they
  should be normalizable modes, so in the gauge theory they correspond
  to states and not to the insertion of sources or couplings
  (c.f.~\cite{Kovtun:2005ev}).} and infalling at the horizon. The
difference is that quasinormal modes decay exponentially in time while
complex momenta describe the decay along the direction of propagation.
The choice of boundary conditions restricts the possible values of
complex frequency or momentum to a discrete set. In the dual gauge
theory we can interpret them as inverse relaxation times $\tau$ or
inverse absorption lengths $\lambda$ of the plasma. The relaxation
time depends on the (real-valued) wave number $k$ whereas the
absorption length depends on the (real-valued) frequency $\omega$,
i.e. $\lambda =\lambda(\omega)$. In the gravity theory we therefore
fix a frequency, impose infalling boundary conditions on the horizon
and search then for a solution of the boundary condition at infinity
in the complexified momentum plane. In this way we can compute the
frequency dependence of the absorption lengths.

Quasinormal modes have been much studied in the context of black holes
in flat spacetime (see \cite{QNreview} for a review) and in the
AdS/CFT correspondence \cite{Horowitz:1999jd, Birmingham:2001pj,
Starinets:2002br, Nunez:2003eq, Kovtun:2005ev, Cardoso:2003cj, 
Konoplya:2003dd}. Complex momenta have
been studied for horizons of compact spatial geometry, where they
correspond to Regge poles \cite{Andersson:1994rk} of the black hole
S-matrix. In the AdS/CFT correspondence, the zero frequency limit of
complex momenta gives the glueball masses of QCD$_3$, as computed in
refs. \cite{Csaki:1998qr, deMelloKoch:1998qs, Brower:2000rp}. While
this work was in progress the interpretation of glueball masses as
correlation lengths was also emphasized in \cite{Bak:2007fk}.

The content of the paper is the following. In section \ref{sec:cmms}
we explain in detail the relation between complex momenta and
absorption lengths. We show that they arise as the poles of the
retarded Green's function and give an argument based on stability
considerations showing that the poles have to lie in the first and
third quadrants of the complex momentum plane. In section
\ref{sec:numerics}, we compute the frequency dependence of the largest
correlation lengths for scalar operators of conformal dimension
$\Delta=4$, global currents, and the transverse and shear channels of
the stress-energy tensor, respectively. We also show the relation with
QCD$_3$ glueball masses. The paper ends with a summary of our results
and some outlook to future possible investigations in section
\ref{sec:conclude}. In appendix \ref{app:effectiveV} we comment on the
form of the effective potentials that arise in rewriting the wave
equations on AdS in the form of a Schr\"odinger equation and in
appendix \ref{app:Heuneq} we show how to avoid the ``false
frequencies'' that arise in the numerical algorithm based on the Heun
equation \cite{Nunez:2003eq}.

\section{Absorption lengths in AdS/CFT} \label{sec:cmms}
In ref. \cite{Son:2002sd} the authors gave a prescription to compute
retarded two-point Green's functions in the context of the AdS/CFT
correspondence with Lorentzian signature. It was emphasized that
retarded propagators correspond to imposing an \textit{infalling}
boundary condition at the horizon for the fields on the gravity side.
On the other hand, infalling boundary conditions are also the
constitutive ingredient for the calculation of the quasinormal
frequencies of black holes in anti de Sitter
space\cite{Horowitz:1999jd}.\footnote{For a more general review of
  quasinormal modes see \cite{QNreview}.} The authors of
\cite{Birmingham:2001pj} observed that the quasinormal frequencies of
BTZ black holes coincide with the poles of the retarded two-point
functions in the dual two dimensional conformal field theory. In
\cite{Nunez:2003eq, Kovtun:2005ev} it was shown that this observation
extends generally to the Lorentzian AdS/CFT correspondence, i.e.
quasinormal frequencies in AdS can be interpreted as the poles of
retarded Green's functions in the dual field theory.

Let us remember the interpretation of the poles of the retarded
Green's function $G_{\rm R}$. The response in the field $\phi$ of the
system under consideration is obtained by the convolution of the
perturbation represented by the source $j(t,\bx)$ with the retarded
Green's function \begin{equation}\phi(t,\bx) := -\int\dd\tau\,\dd\bsxi
  ~G_{\rm R}(t-\tau,\bx-\bsxi) \,j(\tau,\bsxi) ~.  \end{equation}If
one chooses a perturbation localized in time, figuratively speaking
one ``hits" the plasma once at time $t=0$, the perturbation is given
by $j(t,\bx) =\delta(t)\exp(i\bq\bx)$.\footnote{The $x$-dependence is
  that of a plane wave; however a general dependence can be
  constructed by superpositions of plane waves.} Considering the
Fourier transform of the retarded propagator and performing $\tau$ and
$\bsxi$ integrations one arrives at \begin{equation}\phi(t,\bx)
  =-\frac{\,\e^{i\bq\bx}}{2\pi}\int\dd\nu ~\widetilde{G}_{\rm
    R}(\nu,\bq) \,\e^{-i\nu t} ~.  \end{equation}One can now make the
analytical continuation to the complex $\nu$-plane and use Cauchy's
theorem. For $t>0$ we form a closed contour with a semicircle at
infinity on the lower-half $\nu$-plane, whereas for $t<0$ we would
close it in the upper-half $\nu$-plane. One obtains
\begin{equation}\phi(t,\bx) = i\sign(t)\,\e^{i\bq\bx}
  \sum_{\nu_n:\mathrm{poles}} \e^{-i\nu_n t} \,\Res \widetilde{G}_{\rm
    R}(\nu,\bq)\Big|_{\nu=\nu_n} ~, \end{equation}At this point we
assume that the retarded Green's function is analytic in the upper
half of the frequency plane and that its only singularities are single
poles in the lower half plane. This is indeed the analytic structure
that appears in the Lorentzian AdS/CFT correspondence at finite
temperature \cite{Kovtun:2005ev}. In general, the analytic structure
of retarded two-point functions is of course more complicated and
involves also branch cuts. The authors of \cite{Hartnoll:2005ju}
computed the retarded two-point function of $\tr(F^2)$ at weak
coupling and found a tower of branch cuts with branch points located
at $\omega \pm q = -i 4\pi n T$. In this paper we will only consider
the strict large $N$ and strong coupling limit.  Therefore, the
response of the system to a perturbation localized in time is
determined by the sum over the residues of $\widetilde{G}_{\rm R}$ at
the poles. In the holographic gauge theory these poles are
\textit{precisely} the quasinormal frequencies of the perturbation on
AdS space subject to the \textit{infalling} boundary condition.

Instabilities, i.e. exponentially growing modes, appear as quasinormal
frequencies with positive imaginary part. This is consistent with the
interpretation as retarded Green's function, where singularities in
the upper half plane would correspond to tachyonic modes travelling
backwards in time. A typical arrangement of quasinormal frequencies as
they appear in the analysis of small perturbations of asymptotically
AdS black hole spacetimes is depicted in figure \ref{fig:contourQNM}.
\bfig[!htbp] \centering
\includegraphics{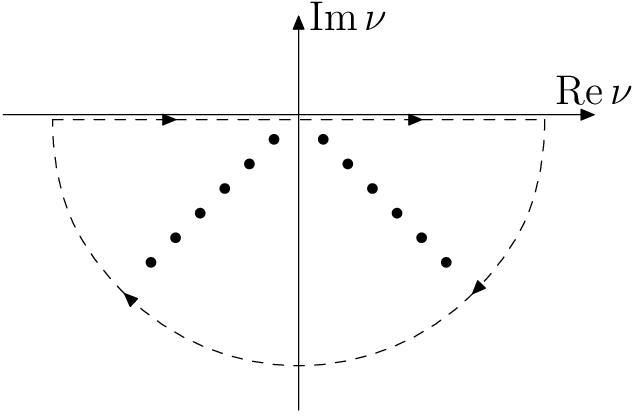}
\caption{\label{fig:contourQNM}The relevant integration contour for
  the poles in the $\nu$-plane. All the poles are in the lower-half
  plane, corresponding to the interpretation of quasinormal modes as
  the poles of a \textit{retarded} Green's function.}
\efig

Let us choose now another kind of perturbation. This time we will pick
a periodic perturbation localized in space, i.e. we switch the roles
of time and one space coordinate and assume a source of the form
$j(t,\bx) =\delta(x)\exp[-i(\omega t -\bk_\perp \bx_\perp)]$. We
compute the effect of such a perturbation again in linear response
theory. Doing the Fourier transform of the retarded propagator and
performing $\bsxi_\perp$ and $\tau$ integrations one finds
\begin{equation} \phi(t,\bx) = -\frac{1}{(2\pi)} \,\e^{-i(\omega
    t-\bk_\perp \bx_\perp)}\int\dd q ~\widetilde{G}_{\rm
    R}(\omega,\bk_\perp, q) \, \,\e^{i q x} ~, \end{equation} This is
the response of the system to a periodic perturbation with frequency
$\omega$ that is localized in the $x$-direction and has the form of a
plane wave in the perpendicular directions $\bx_\perp$. We have
assumed that the perturbation has started far in the past such that
all transient oscillations have already vanished and the system has
reached a stationary state. In the following we will also assume that
the perturbation is not further modulated in the
$\bx_\perp$-directions, i.e. we set $\bk_\perp = 0$.  Now one can use
again Cauchy's theorem, closing in the upper or lower-half planes for
$x>0$ and $x<0$, respectively. The result is
\begin{equation}
  \phi(t,x) = - i \,\sign(x)\,\e^{-i\omega t} \sum_{q_n:\mathrm{poles}}
  \e^{iq_n x} \,\Res \widetilde{G}_{\rm R}(\omega,q)\Big|_{q=q_n} ~,
\end{equation}
Again we have assumed that the only singularities of $G_{\rm
  R}(\omega,\bq)$ are poles in the complexified momentum plane. By
symmetry considerations ($x\rightarrow -x$) it is clear that if $q
=q_n =q^R_n + i q^I_n$ is a pole then also $q=-q_n$ has to be a pole.
We would like the poles in the upper-half to lie in the first quadrant
and those in the lower-half in the third quadrant. With such and
arrangement of poles the perturbation is creating damped waves moving
to the right for $x>0$ and to the left for $x<0$. The waves propagate
away from the origin of the perturbation at $x=0$ and are
exponentially decaying with the distance from the perturbation. In
subsection \ref{ssec:locCMMs} we prove that for the holographic
retarded two-point functions the poles do indeed fall into the first
and third quadrants of the complex momentum plane. A typical setup
with the corresponding integration contours is depicted in figure
\ref{fig:contourCMM}.
\bfig[!htbp] \centering
\includegraphics{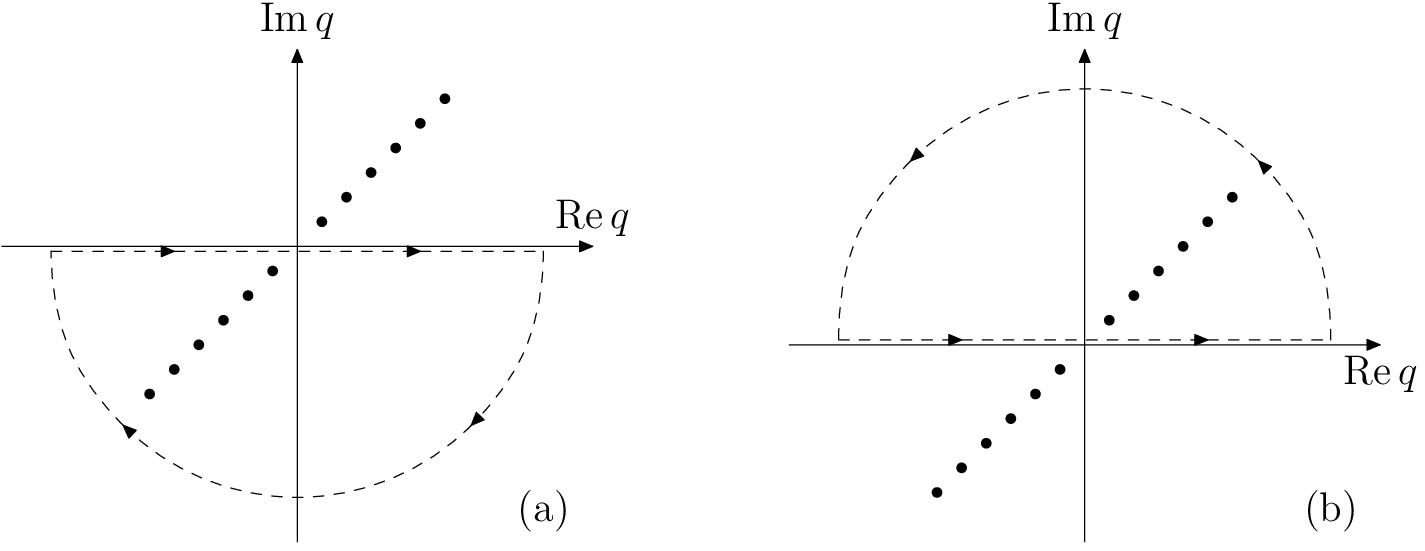}
\caption{\label{fig:contourCMM}The relevant integration contours for
  the poles in the complexified momentum-plane. Figure (a) shows the
  contour for $x<0$, and figure (b) shows the contour for $x>0$. In
  order to obtain exponentially decaying waves travelling away from
  the origin of the perturbation it is necessary that the poles lie in
  the 1st and 3rd quadrants.}
\efig
The imaginary part of the complex wave number can be interpreted as
the inverse of an \textit{absorption length}. For a given complex
momentum pole $q_n$ the right-moving wave has the form $\e^{-i(\omega
  t - q_n^R x)} \,\e^{-q_n^I x}$. The amplitude of the wave has
decayed to a factor of $1/e$ at a distance of $\lambda_{n} = 1/q_n^I$.

In the following we will be interested in computing these absorption
lengths and their frequency dependence in the holographic dual of the
$\Nfour$ supersymmetric gauge theory in the plasma phase. We will do
this for different kinds of perturbations corresponding to certain
gauge-invariant operators. In the gravity side we have to solve wave
equations with purely infalling boundary conditions at the horizon
just as in the calculation of quasinormal modes. At the boundary of
AdS we have to specify the same boundary conditions that have been
described in refs. \cite{Nunez:2003eq, Kovtun:2005ev} for the
quasinormal modes. The imaginary part of these complex momentum wave
numbers give absorption lengths characteristic of the black hole.
After having traveled a distance $\lambda_n$ a part of the wave has
fallen into the black hole such that the amplitude is diminished by a
factor of $e^{-1}$. In the gauge theory dual the inverse of the
imaginary part of the momenta can be seen as the absorption lengths
for perturbations of the plasma by sources corresponding to
gauge-invariant operators.

Thus, we see that in both cases -- relaxation times and absorption
lengths -- the gravity waves are subject to the infalling condition at
the horizon. The question is simply which parameter of the retarded
Green's function is analytically continued to complex values, either
the frequency or the momentum. To compute these complex momentum wave
numbers one can therefore follow the same strategy that is used for
the calculation of quasinormal frequencies, but fixing the frequency
$\omega$ to be real-valued instead of the momentum $q$.

This switch of roles is particularly clear in the case of the $\rm
AdS_3/CFT_2$ correspondence, where the exact retarded Green's
functions can be calculated in both sides and seen to match
\cite{Birmingham:2001pj}. Let us consider the case of a field with
conformal dimension $\Delta = 2$. Then the retarded two-point function
is
\begin{equation}
  G_{\rm R}^{\rm (2d)} (\omega, q) =\frac{\omega^2 -q^2}{4\pi^2} \left[
    \psi\left( 1 -i\,\frac{\omega-q}{4\pi T} \right) +\psi\left( 1
      -i\,\frac{\omega+q}{4\pi T} \right) \right] ~.
\end{equation}
The poles of the $\psi$ function determine the quasinormal frequencies
$\omega_n = \pm q - i 4\pi T (n+1)$. For each quasinormal mode the
dispersion relation $\omega_n = \omega_n(q)$ is linear. Because of
this linearity the poles can also be interpreted in a different way by
writing
\begin{equation}
  q_n = \pm [ \omega + i 4 \pi T (n+1) ] ~,\quad n\in\bbN ~,
\end{equation}
where we see explicitly that the complex momentum modes lie in the
first and third quadrants for the right- and left-movers respectively.

In higher dimensions the dispersion relations for the quasinormal
frequencies are not linear and can be computed only numerically. At zero momentum, the position of large frequencies in the complex momentum plane has been estimated using semiclassical methods \cite{Natario:2004jd}, it would be interesting to extend those analysis to non-zero momentum.
. Since
the dispersion relation for the quasinormal modes is known only
numerically we also have to resort to numerical methods to find the
complex wave numbers and absorption lengths. The only exception is
given by the hydrodynamic modes that appear for small frequency and
wave numbers \cite{Policastro:2002se, Policastro:2002tn}. We will see
that our numerical results are in agreement with the analytic
dispersion relations of the hydrodynamic modes.

\subsection{Stability analysis} \label{ssec:locCMMs} We will now
perform a stability analysis analogous to the one for quasinormal
modes in \cite{Horowitz:1999jd}. We will see that the complex momentum
wave numbers indeed lie in the first and third quadrants of the
complex $q$-plane for positive frequencies. Note that a pole in the
second or fourth quadrant would allow to construct outgoing waves that
are exponentially growing with the distance from the perturbation. For
the stability of the system under the perturbation the absence of such
poles is therefore crucial.

The time and space dependence of the field is given by simple
exponentials
\[
\phi(t,x) \sim \e^{-i\omega t} \,\e^{iqx} ~,
\]
where $q:=q^R +iq^I$. We will distinguish between the cases $x>0$ and
$x<0$.

For $x>0$ stability demands an exponentially decaying wave and
therefore $\sign q^I = +1$. We further demand that the wave is
outgoing from the origin of the perturbation which demands $\sign
\omega = \sign q^R$. Taking these two facts together amounts to the
condition \begin{equation}\label{eq:qdrntconds} \sign\left(
    \frac{\omega}{q^R} \right) = \sign q^I ~.  \end{equation} Doing
the same analysis for $x<0$, one finds that the perturbation moves
away to the left if $\sign \omega = -\sign q^R$, whereas the stability
condition is now $\sign q^I =-1$.  This again amounts to eq.
\erf{eq:qdrntconds}.

We want to prove now that in the gravity dual the complex momentum
modes of the black hole follow indeed the rule given by equation
\erf{eq:qdrntconds}. We consider a minimally coupled scalar $\Phi$ in
\adsfive with mass $m$. The line element of the AdS black hole with
planar horizon is \begin{equation}\label{eq:AdSmetric} \dd s^2_{\rm
    AdS} = \frac{r^2}{R^2} \left( -f(r)\dd t^2 +\dd\bx^2 \right)
  +\frac{R^2}{r^2 f(r)} \,\dd r^2 ~, \end{equation}where
$f(r)=1-(r_{\rm H}/r)^4$. The temperature is given through $r_{\rm H}
= \pi R^2 T$. We will use in the following the coordinate $z=r_{\rm
  H}/r$ and rescale time and space coordinates by $r_{\rm H} R^{-2}
(t,\bx) \mapsto (t,\bx) $. The boundary is now located at $z=0$ and
the horizon at $z=1$. The equation of motion for a minimally coupled
scalar $\Phi(t,z,x) = \exp(-i\omega t+iqx) \,\Phi(z)$ of mass $m$ is
\begin{equation}\Phi'' + \left( \frac{f'(z)}{f(z)} -\frac{3}{z}
  \right) \Phi' + \left( \frac{\omega^2}{f(z)^2} -\frac{q^2}{f(z)}
    -\frac{m^2}{z^2 f(z)} \right) \Phi(z) =0 ~.  \end{equation}

We can further split $\Phi(z) =\sigma(z) \,y(z)$, in order to find an
equation for $y(z)$ that is Schr\"odinger-like in a `tortoise' $z_*$
coordinate defined through \begin{equation}\dd z_* =\frac{\dd z}{f(z)}
  \quad \Rightarrow \quad \Big( \partial^2_{z_*} +\omega^2 -V(z_*)
  \Big) y(z_*) =0 ~, \end{equation} provided that $\sigma(z)$ fulfils
\be \frac{\sigma'(z)}{\sigma(z)} =\frac{3}{2z} ~.  \end{equation}In
the $z$ coordinate the Schr\"odinger potential reads
\begin{equation}V(z) = \frac{f(z)}{4z^2} \Big( 15 +4m^2 +4q^2z^2 +9z^4
  \Big) :=V_0(z) +\Re(q^2)\,f(z)+i\Im(q^2) f(z) ~, \end{equation}where
we have separated it into its real and imaginary parts. In the $z_*$
coordinate the horizon lies at $z_*\to +\infty$ and the potential
vanishes there, so the wavefunction can be described as a
superposition of plane waves. The infalling boundary condition
corresponds to setting $y(z_*) =\e^{i\omega z_*} \,\chi(z_*)$ with
$\chi(+\infty)=\,{\rm const}$.  Thus we find
\be\label{eq:schroedinger} \Big( \partial^2_{z_*}
+2i\omega\,\partial_{z_*} -V_0(z_*)
-\Re(q^2)\,f(z_*)-i\Im(q^2)\,f(z_*) \Big) \,\chi(z_*) =0 ~.
\end{equation}If we multiply by the conjugate $\overline{\chi}(z_*)$
and pick out the imaginary part of the equation we obtain
\begin{equation}\label{eq:qdrnteq} -\frac{i}{2}
  \,(\overline{\chi}\partial^2_{z_*}\chi
  -\chi\partial^2_{z_*}\overline{\chi})
  +\omega\,\partial_{z_*}|\chi|^2 -\Im(q^2)f(z_*)|\chi|^2 =0 ~,
\end{equation}Now we integrate this equation between the boundary
($z_*=z_*^b$) and the horizon ($z_*=+\infty$).  Upon a partial
integration the derivative terms cancel each other: $\chi(z_*)$
vanishes at the boundary due to the Dirichlet boundary condition we
impose there and at the horizon the derivative vanishes
$\partial_{z_*}\chi(+\infty) =0$. The remaining terms in equation
\erf{eq:qdrnteq} amount to \begin{equation}
\label{eq:stability.im} \omega
\,|\chi(z=1)|^2 =\Im(q^2) \int_{z_*^b}^\infty \dd z_*
\,f(z_*)|\chi(z_*)|^2 =\Im(q^2) \int_0^1 \dd z \,|\chi(z)|^2 ~, \end{equation}

The integral on the right hand side is positive definite, which then
implies that $\sign\omega = \sign\Im(q^2) =\sign(q^R q^I)$ which is
precisely the stability condition \erf{eq:qdrntconds}.

There is a further stability condition involving the properties of the
potential. When $\omega=0$, we have the condition that $\Im(q^2)=0$,
so either $q^R=0$ or $q^I=0$. Consider now the real part of the
Schr\"odinger equation \erf{eq:schroedinger}. After multiplying by
$\overline{\chi}(z_*)$ and integrating between the boundary and the
horizon we find \begin{equation}\label{eq:stability.re} \int_{z_*^b}^\infty \dd z_*
\left( |\partial_{z_*}\chi(z_*) + i \omega \chi(z_*)|^2
  +(V_0(z_*)-\omega^2) |\chi(z_*)|^2 \right)= -\Re(q^2)\int_0^1 \dd z
\,|\chi(z)|^2 ~, \end{equation}
Clearly, if $V_0(z_*)\geq 0$ between the boundary
and the horizon, then, at $\omega=0$, $\Re(q^2)<0$ and we will have
$q^R=0$ and $q^I\neq 0$. On the other hand, if the potential is
negative on some region, then there could be solutions with
$\Re(q^2)>0$ or equivalently $q^R\neq 0$ and $q^I=0$. Considering
four-dimensional Minkowski slices of AdS$_5$, these modes can be
regarded as tachyonic instabilities of negative mass squared
$m^2=-(q^R)^2$. Notice that with our choice the boundary conditions $\sim
e^{i\omega z_*}$ and fixing $\omega$ to be real, this condition actually refers
to the presence of 'negative energy' modes in the scattering spectrum, so only
when the potential is negative at the horizon this kind of instabilities could
appear. Other instabilities associated to the presence of bound states could be
present, see the appendix \ref{app:effectiveV} for a discussion.

If $\Re(q^2)>0$ instabilities are present in the bulk theory, the gauge
correlation functions associated to the dual operators will show an
oscillatory behavior at large separations, as opposed to vanishing,
indicating that the plasma is actually out of equilibrium. From the
point of view of the effective three-dimensional theory, instabilities
will appear as tachyonic states in the spectrum.

\section{Absorption lengths: numerical analysis}  \label{sec:numerics}
\subsection{Scalar operators}\label{sec:scalar}
As a first example we want to compute the absorption lengths of a
scalar operator ${\cal O}(t,\bx)$ of conformal dimension $\Delta =4$.
We choose this particular case because it is the simplest setup we can
use to illustrate the method, since the dual supergravity field
corresponds to a minimally coupled, massless scalar. A possible
example is given by ${\cal O}={\rm tr} F^2$, that maps to the dilaton
in the holographic dual.

Consider the retarded two-point correlation function in the theory at
temperature $T$ \begin{equation}G_{\rm R} (t-t',\bx-\by)=-i \,\theta(t-t')
\,\vev{
  [\cO(t,\bx),\cO(t',\by)] } ~.  \end{equation}At large distances $|\bx -\by|
\gg T^{-1}$, the Green's function decays exponentially due to thermal
screening. As we have explained, this behavior is determined by a set
of discrete lengths that in linear response theory describe the {\em
  absorption} of out-of-equilibrium perturbations. For the theory in
equilibrium they are identified with {\em correlation lengths} in the
plasma, as proposed in ref. \cite{Bak:2007fk}. The squared inverse of
the zero-frequency correlation lengths can also be regarded as the
glueball masses\,\footnote{In this particular example we are
  considering $J^{PC}=0^{++}$ glueballs.} of a three-dimensional
effective theory in the high-temperature limit \cite{Csaki:1998qr,
  deMelloKoch:1998qs, Brower:2000rp}. Via AdS/CFT correspondence we
can reduce this complicated non-perturbative problem in the gauge
theory to finding the complex momenta that allow the dilaton
fluctuations to obey infalling boundary conditions on the horizon and
Dirichlet ones on the boundary. In this example, and in the other
cases we consider in this paper, the equations of motion can be
reduced to Heun equations that we can solve using semi-analytic
methods.

The equation of motion for this field was already derived in
subsection \ref{ssec:locCMMs}. Throughout the paper we use
dimensionless frequency and momentum. In order to recover the
dimensionful quantities it is enough to make the substitution
$(\omega,\bq) \mapsto \pi T\,(\omega,\bq)$. Changing coordinates from
$z$ to $x=1-z^2$, the equation now reads \begin{equation}\label{eq:scalarEOM}
\Phi'' +\frac{1+(1-x)^2}{x(1-x)(2-x)}\,\Phi' +\left(
  \frac{\omega^2}{4x^2(1-x)(2-x)^2} -\frac{q^2}{4x(1-x)(2-x)}
\right)\,\Phi(x) =0 ~.  \end{equation}

This equation has four regular singular points at $x=0,1,2,\infty$,
with characteristic exponents
\[
\{0;-i\omega/4,+i\omega/4\} ~,\quad \{1;0,2\} ~,\quad
\{2;-\omega/4,+\omega/4\} ~,\quad \{\infty;0,0\} ~.
\]
Therefore, we can transform it into a Heun equation and we can follow
the analysis described in \cite{Starinets:2002br}. To compute the
complex wave numbers we simply have to analytically continue the
momentum instead of the frequency. It is interesting to observe that
none of the characteristic exponents at the singular points depend on
the momenta. We factorize $\Phi(x)$ as \begin{equation}\Phi(x) =x^{-i\omega/4}
\,(1-x)^2 \,(2-x)^{-\omega/4} \,y(x) ~, \end{equation}which allows us to write
the equation of motion in the standard form of a Heun equation for
$y(z)$ \begin{equation}y''(x) +\left( \frac{\gamma}{x} +\frac{\delta}{x-1}
  +\frac{\epsilon}{x-2} \right)y'(x) +\frac{\alpha\beta
  x-Q}{x(x-1)(x-2)}\,y(x) =0 ~, \end{equation}with parameters
\begin{eqnarray}\label{eq:scalarparameters}
\alpha &=& \beta=2 -\frac{\omega}{4}(1+i) ~,\quad Q =4 +\frac{q^2}{4}
-(1+7i)\frac{\omega}{4} -(2-i)\frac{\omega^2}{8} ~, \\
\gamma &=& 1-i\frac{\omega}{2} ~,\quad \delta=3 ~,\quad \epsilon
=1-\frac{\omega}{2} ~.  \end{eqnarray} All the other perturbations we will
consider in this paper can be transformed to Heun equations in a
similar way. In \cite{Nunez:2003eq} the same perturbations have been
studied in order to solve for the quasinormal modes. The solution
$y_0(x)$ corresponding to the infalling boundary conditions at the
horizon $x=0$ is a linear combination of local solutions $y_{1,A}$,
$y_{1,B}$ at $x=1$
\begin{equation}
  y_0(x) = A\,y_{1,A}(x) + B\,y_{1,B}(x) ~,
\end{equation}
where $y_{1,A}$ is analytic at $x=1$. The retarded Green's function
turns out to be proportional to
\begin{equation}
  G_{\rm R} \propto \frac{A}{B} ~,
\end{equation}
and the poles correspond therefore to the solutions with $B=0$, i.e.
solutions that are analytic in the interval $x\in[0,1]$. These
boundary conditions determine a discrete set of complex momentum
eigenvalues if we fix the frequency $\omega$ to real values.

We can find local solutions using the Frobenius method close to the
singularities. In the cases under consideration, a solution with
boundary condition $y(0)=\,{\rm const.}$ will be a superposition of
solutions with exponents $1-\delta \leq 0$ and $0$ close to the AdS
boundary ($x=1$). The boundary condition $y(1)=\,{\rm const.}$ can be
satisfied only for a discrete set of frequencies $\omega$ or momenta
$q$. These values can be computed imposing matching conditions at some
intermediate point for the Frobenius series although we will need a
large number of terms and the convergence gets worse for higher
frequencies. There is an alternative method based on the improved
convergence of the solutions. Normal solutions are convergent for
$|x|<1$, but for some values of the parameters the solutions can
converge for $|x|<2$.  This condition of extended convergence boils
down to a transcendental equation for the frequencies or momenta in
the form of a continued fraction (see
\cite{Leaver:1985ax,Starinets:2002br} for more details) using
Pincherle's theorem on the existence of minimal solutions to three
term recursion relations.

The coefficients of the Frobenius series at $x=0$ should satisfy the
recursion relation
\begin{equation}\label{eq:heun.recursion}
  a_{n+2} +A_n(\omega,q) \,a_{n+1} +B_n(\omega,q) \,a_n=0 ~,\quad n\geq 0 ~,
\end{equation}
where 
\begin{eqnarray}\label{eq:ttrr}
A_n(\omega,q) &=&
-\frac{(n+1)(2\delta+\epsilon+3(n+\gamma))+Q}{2(n+2)(n+1+\gamma)} ~, \\
B_n(\omega,q) &=& \frac{(n+\alpha)(n+\beta)}{2(n+2)(n+1+\gamma)} ~,
\end{eqnarray}
 and $a_0=1, ~a_1=Q/2\gamma$. Then, using the recursive definition
\begin{equation}\label{eq:continuedfraction}
  r_n =\frac{a_{n+1}}{a_n} = -\frac{B_n(\omega,q)}{A_n(\omega,q)+r_{n+1}} ~.
\end{equation}
Pincherle's theorem states that a minimal solution to the three term
recursion relation \erf{eq:ttrr} exists if and only if the continued
fraction on the right hand side in \erf{eq:continuedfraction}
converges. Moreover, in this case it converges precisely to
$a_{n+1}/a_n $. In \cite{Starinets:2002br, Nunez:2003eq} it was
pointed out that the minimal solution corresponds precisely to a
solution of the Heun equation that is analytic at $x=1$ therefore
fulfilling the correct boundary conditions. Choosing $n=0$ we find
\begin{equation}\label{eq:r0}
  r_0 =\frac{Q}{2\gamma} ~.
\end{equation}
and computing $r_0$ recursively gives a transcendental equation for
$q$.  Using this formula, we can compute numerically the complex
momentum modes with high precision. In order to do that, we cut the
fraction at a large value $n=n_*=100$ and use the asymptotic value
$r_n=1/2-(2+\omega)/4n_*$. It is important to realize that Pincherle's
theorem applies only if we are dealing with genuine three term
recursion relations. Sometimes it can happen that the recursion
relation involves three terms only from a certain value of $n=n_1$ on.
This happens for example if either $\alpha=0$ or $\beta=0$ when
$B_0=0$. In such a case one has to use \erf{eq:continuedfraction} with
$n=n_1$. We will see that we are faced with this in the cases of the
longitudinal vector field perturbations and of the shear mode
perturbations at $\omega=0$. Since $\alpha=0$ in both cases it is
sufficient to take $n_1=1$ and use
\begin{equation}\label{eq:r1}
  r_1 = \frac{Q^2 + 3 Q \gamma - 2\alpha \beta\gamma + 2 Q\delta 
    + Q\epsilon}{4 Q + 4Q\gamma}~,
\end{equation}
instead of \erf{eq:r0}.

\bfig[!htbp] \centering
\includegraphics[scale=0.9]{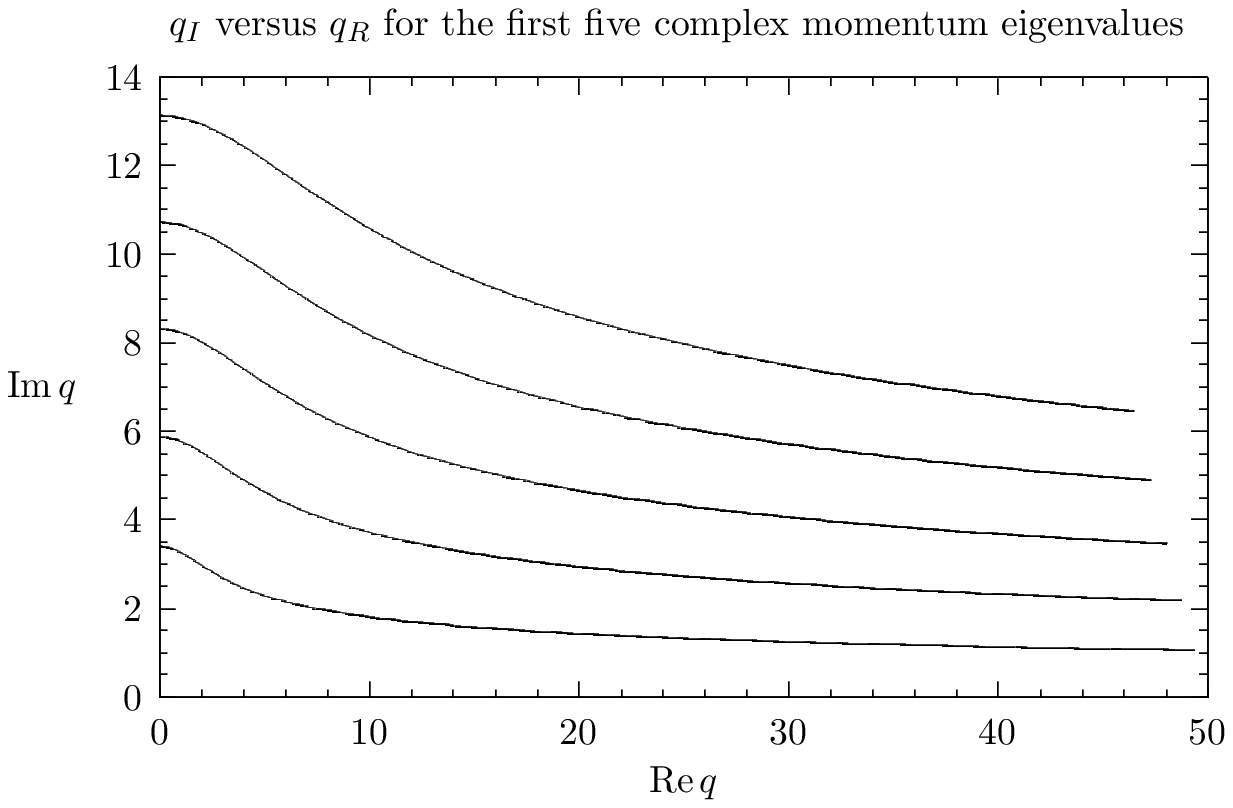}
\caption{\label{fig:scalarqIqR} We have traced the locations of the
  five lowest momentum eigenvalues in the complex $q$-plane for
  different frequencies as a function of the frequency out to
  $\omega=50$. The momentum eigenvalues vary continuously with the
  frequency and lie on the analogues of Regge trajectories.}
\efig
\TABLE[p]{
  \begin{tabular}{|c|r|} \hline\hline $n$ & $q_n^2$ \hspace*{3.5ex} \\
    \hline\hline \texttt{1} & $-11.5877$ \\ \hline \texttt{2} &
    $-34.5270$ \\ \hline \texttt{3} & $-68.9750$ \\ \hline \texttt{4}
    & $-114.9104$ \\ \hline \texttt{5} & $-172.3312$ \\ \hline
    \texttt{6} & $-241.2366$ \\ \hline \texttt{7} & $-321.6265$ \\
    \hline \texttt{8} & $-413.5009$ \\ \hline \texttt{9} & $-516.8597$
    \\ \hline \texttt{10} & $-631.7028$ \\ \hline
  \end{tabular}
  \caption{\label{tab:gluebQCD3}The first ten glueballs of the scalar
    mode.}}

We have numerically computed the complex momentum eigenvalues using
this method. The results for the scalar field perturbations are shown
in figure \ref{fig:scalarqRw}. The real and imaginary parts of the
five lowest complex momentum are plotted as a function of the
frequency. The real parts start out at $q^R =0$ for zero frequency.
The imaginary parts start out at a finite value at $\omega=0$, develop
a shoulder that is more pronounced for the higher modes and then fall
off rather fast until they enter a regime of slow decrease for large
frequencies. Numerically we found that the lowest mode becomes almost
constant at large frequencies with $q^I \approx 0.83$ at $\omega=100$.
Also the higher modes flatten out for high frequencies. As expected,
higher frequencies can penetrate farther into the plasma. It is an
interesting question if the plasma becomes transparent for some high
but finite frequency, if transparency is reached only in the limit
$\omega=\infty$ or if the absorption length stays finite.
Unfortunately our algorithm does not allow us to explore this
asymptotic regime. We can speculate however using the underlying
conformal invariance of the $\Nfour$ theory. Since for high
frequencies the temperature is less and less important we expect that
the absorption length diverges as $\omega \to \infty$, i.e.
$q^I(\omega=\infty) = 0$. A finite absorption length would point to an
underlying scale in the theory. On the other hand, if the plasma were
to become transparent at some finite value of $\omega$, we would
expect that to happen at a scale that is set by the temperature.
However, our numerical results show finite absorption lengths for much
higher frequencies.

\subsubsection{Glueball masses} \label{ssec:glueballs}
Of particular interest are the absorption lengths in the static limit
$\omega\to 0$. In this case we will refer to the absorption length as
the screening length. The equation \erf{eq:scalarEOM} with $\omega=0$
has been studied before in \cite{Csaki:1998qr,deMelloKoch:1998qs}.
There the interpretation of the eigenvalues in the momentum with
$q^2<0$ was as masses of glueballs in the three dimensional theory
that is obtained by reduction on the thermal circle in the Euclidean
section of the AdS black hole. The glueball masses can be calculated
as the discrete eigenvalues $M_n^2=-q_n^2$. Our numerical results at
$\omega = 0$ for the first ten modes are compiled in table
\ref{tab:gluebQCD3} and are in good agreement with results given by
refs. \cite{Csaki:1998qr, deMelloKoch:1998qs}.

It is important to see if the eigenfunctions correspond to the wave
functions of the glueballs too. In \cite{Brower:2000rp} the authors
observe that for all the glueball masses the correct boundary
conditions correspond to demanding analyticity of the wave function at
the horizon and the boundary. These are precisely the same boundary
conditions that emerge in our case at $\omega = 0$. Therefore the
screening lengths for static fields corresponds precisely to the
glueball masses computed earlier in \cite{Csaki:1998qr,
  deMelloKoch:1998qs, Brower:2000rp}.

\subsection{Global currents}  \label{sec:vector}
In the $\Nfour$ theory, the global currents associated to R-charges
map to mixed components of the ${\rm S}^5$ and AdS$_5$ metrics, that
can be seen as graviphotons after dimensional reduction to AdS$_5$. In
general, any global symmetry in the field theory will map to a local
gauge symmetry in the holographic dual. Then, to find the poles of the
retarded Green functions in the plasma
\begin{equation}G_{\mu\nu}(t-t',\bx-\by)= -i \,\theta(t-t') \,\vev{
    [J_\mu(t,\bx),J_\nu(t',\by)] } ~, \end{equation}we have to compute
the complex momentum eigenvalues for vector fields in the AdS black
hole.\footnote{We are assuming that the total charge in the
  equilibrium state vanishes, so there are no chemical potentials.} We
will see that there are two decoupled sectors, corresponding to
transverse and longitudinal channels. The reason is that temperature
breaks four-dimensional Lorentz symmetry to three-dimensional
rotational symmetry.  In the glueball language, the zero-frequency
masses correspond to $J^{PC}=1^{--}$ and $0^{-+}$ states. However the
states arising from the vector fields are also charged under the
global R-symmetry and therefore do not form part of the superselection
sector that constitutes QCD$_3$.  For simplicity we will refer to
these states also as glueballs.  The longitudinal channel is special
because it also describes the diffusion of the conserved charge
through the plasma, that does not appear as a glueball state in the
three-dimensional theory because the residue of the diffusion mode
vanishes in the zero-frequency limit. We will show that the diffusion
pole is also captured by complex momentum eigenvalues.

We can compute the complex momentum eigenvalues corresponding to a
vector field in the AdS-Schwarzschild background in an analogous way
to the scalar field case. The equations of motion for such a field are
given by the Maxwell equations \begin{equation}\frac{1}{\sqrt{-g}}
  \partial_\nu \left[ \sqrt{-g} \,g^{\mu \rho} g^{\nu \sigma} F_{\rho
      \sigma} \right] =0 ~, \end{equation}where
$F_{\mu\nu}=\partial_\mu A_\nu - \partial_\nu A_\mu$.  We can choose
the $A_z=0$ gauge in the metric \erf{eq:AdSmetric} and expand in plane
waves $A_\mu$. Separating the vector field in longitudinal and
transverse components, $A_L(z)=\frac{\bq}{q} \cdot \bA(z)$ and
$\bq\cdot \bA_T(z)=0$, the equations of motion are
\begin{eqnarray}
  \label{eq:vectorconstraint}
  0 &=& \omega A'_0(z) + f(z) q A'_L(z) ~, \\
  0 &=& A''_0(z) - \frac{1}{z} A'_0(z) -\frac{1}{f(z)} \Big( \omega q A_L(z)+q^2
  A_0(z)\Big) ~, \\
  0 &=& A''_L(z) +\frac{z}{f(z)} \left(\frac{f(z)}{z}\right)' A'_L(z)
  +\frac{1}{f(z)^2}\Big(\omega q A_0(z)+\omega^2 A_L(z)\Big) ~, \\
  0 &=& \bA''_T(z) + \frac{z}{f(z)} \left(\frac{f(z)}{z}\right)'
  \bA'_T(z)
  +\left(\frac{\omega^2}{f(z)^2}-\frac{q^2}{f(z)}\right)\bA_T(z) ~.
\end{eqnarray}
The first three equations are not independent so we can use the first
one to write decoupled equations for $A'_0(z)$ and $A'_L(z)$.  Notice
that there is a choice of gauge invariant variables $E_L=q A_0+\omega
A_L$ and $\bE_T=\omega \bA_T$ that describe the diffusive and
transverse channel respectively \cite{Kovtun:2005ev}. However, the
spectrum of complex momentum values (equivalently of quasinormal
modes) is gauge invariant, so it should not matter if we choose to
work with gauge components, that obey simpler equations. Since the
invariant quantity is $E_L$, this means that $A_0$ and $A_L$ should
have the same spectrum, as the constraint (\ref{eq:vectorconstraint})
points out.

Then, the relevant equations for the temporal, longitudinal and
transverse components of the vector field read
\begin{eqnarray} \label{eq:vector1} 0 &=&
  V''_0+\left(\frac{f'}{f}-\frac{1}{z}\right)V'_0 +
  \left(\frac{\omega^2}{f^2}-\frac{q^2}{f}-\frac{f'}{zf}+\frac{1}{z^2}
  \right)V_0(z) ~, \\
  \label{eq:vector2}
  0 &=& V''_L+\left(3\frac{f'}{f}-\frac{1}{z}\right)V'_L +
  \left(\frac{\omega^2}{f^2}-\frac{q^2}{f}+\left(\frac{f'}{f}\right)^2
    +\frac{f''}{f} -2\frac{f'}{zf} +\frac{1}{z^2} \right)V_L(z) ~, \\
  0 &=& \bA''_T + \left(\frac{f'}{f}-\frac{1}{z}\right)\bA'_T +
  \left(\frac{\omega^2}{f^2}-\frac{q^2}{f}\right)\bA_T(z) ~.  
\end{eqnarray}
where we define $V_0(z)=A'_0(z)$, $V_L(z)=A'_L(z)$. In the $x=1-z^2$
coordinate and for a suitable factorization of each component, the
equations above can be written as Heun equations:
\paragraph*{Temporal.}  The critical exponents at the singularities
are
\[
\left(0;-i\frac{\omega}{4},i\frac{\omega}{4}\right) ~,\quad
\left(1;\frac{1}{2},\frac 1 2\right) ~,\quad
\left(2;-\frac{\omega}{4},\frac{\omega}{4}\right) ~,\quad
\left(\infty;-\frac 1 2,\frac 3 2\right) ~.
\]
\begin{equation}V_0(x)=x^{-i \omega/4} (x-1)^{1/2} (x-2)^{-\omega/4}
  y(x) ~.
\end{equation}

\paragraph*{Longitudinal.} The critical exponents at the singularities
are
\[
\left(0;-1-i\frac{\omega}{4},-1+i\frac{\omega}{4}\right) ~,~~
\left(1;\frac 1 2,\frac 1 2\right) ~,~~
\left(2;-1-\frac{\omega}{4},-1+\frac{\omega}{4}\right) ~,~~
\left(\infty; \frac 5 2-\sqrt{5},\frac 5 2+\sqrt{5}\right) ~.
\]
\begin{equation}V_L(x)=x^{-1-i \omega/4} (x-1)^{1/2}
  (x-2)^{-1-\omega/4} y(x) ~.
\end{equation}In both cases we find the same parameters for the Heun equation
\bea\label{eq:longitudinalparameters}
\alpha&=& -\frac{\omega}{4}(1+i)~,\quad \beta= 2-
\left(\frac{\omega}{4}(1+i)\right) ~,\quad Q =\frac{q^2}{4}
-(1+3i)\frac{\omega}{4} -(2-i)\frac{\omega^2}{8} ~,  \nonumber \\
\gamma &=& 1-i\frac{\omega}{2} ~,\quad \delta=1 ~,\quad \epsilon
=1-\frac{\omega}{2} ~.  \eea Notice that the boundary conditions for
$V_L(x)$ are not infalling ones. They are determined by the constraint
\ref{eq:vectorconstraint}.
\paragraph*{Transverse.} The critical exponents at the singularities
are
\[
\left(0;-i\frac{\omega}{4},i\frac{\omega}{4}\right) ~,\quad
\left(1;0,1\right) ~,\quad
\left(2;-\frac{\omega}{4},\frac{\omega}{4}\right) ~,\quad
\left(\infty;0,1\right) ~.
\]
\begin{equation}A_T(x)=x^{-i \omega/4}(x-1)(x-2)^{-\omega/4} y(x) ~.
\end{equation}
\begin{eqnarray}
  \alpha&=& 1-\frac{\omega}{4}(1+i) ~,\quad \beta =2-
  \left(\frac{\omega}{4}(1+i)\right)~,\quad Q =\frac{q^2}{4}
  +2-(1+5i)\frac{\omega}{4} -(2-i)\frac{\omega^2}{8} ~,  \nonumber \\
  \gamma &=& 1-i\frac{\omega}{2} ~,\quad \delta=2 ~,\quad \epsilon
  =1-\frac{\omega}{2} ~.  
\end{eqnarray}

As we had anticipated, the temporal and longitudinal components have
the same spectrum, since they obey the same Heun equation, although
this was not evident in equations \erf{eq:vector1} and
\erf{eq:vector2}.

The results are shown in figures \ref{fig:longwthomega} and
\ref{fig:transvwthomega}. The real and imaginary parts of the five
lowest complex momentum are plotted as a function of the frequency.
The behavior is similar to the one found for the scalar operator. The
imaginary parts start out at a finite value at $\omega=0$, develop a
shoulder that is more pronounced for the higher modes and then fall
off rather fast until they enter a regime of slow decrease for large
frequencies. The real parts start out the $q^R =0$ for zero frequency.

So far, we have described the absorption of R-current excitations in
the plasma. However, a conserved global charge cannot be dissipated,
it is spread out by the slow process of diffusion. This is described
in the hydrodynamic regime $\omega, q \ll T$ by a diffusion pole
\cite{Policastro:2002se} (units restored)
\begin{equation}
  \omega= -i \,\frac{q^2}{2 \pi T} ~.
\end{equation}
In our analysis of complex wave numbers we are able to see numerically
this mode ($q= (1+i)\sqrt{\omega}$ with our conventions) that fits
nicely with the analytic prediction in the hydrodynamical regime
Fig.~\ref{fig:hydrovec}.

\subsubsection{Glueball masses}\label{ssec:glueballsVEC}
In the zero frequency limit, the absorption lengths can be interpreted
as the inverse glueball masses of an effective three-dimensional
theory. Note however that these states do not lie in the
superselection sector that constitutes the holographic dual of
QCD$_3$! For the longitudinal channel we have to take into account
that $\alpha(\omega=0)=0$ so we have to use the modified recursion
relation starting at $n=1$ \erf{eq:r1}. It turns out that the glueball
masses of the longitudinal channel coincide with the ones found for
the scalar operator, table~\ref{tab:gluebQCD3}.  Indeed, at $\omega=0$
we can transform the Heun equation with
parameters~\erf{eq:scalarparameters} into the Heun equation with
parameters~\erf{eq:longitudinalparameters}. First, make the coordinate
transformation $x\to \frac{x-2}{x-1}$, that shuffles the singular
points $2\leftrightarrow 0$, $1\leftrightarrow \infty$. Then, the
redefinition $y(x)\to (x-1)^2 y(x)$ (explained in the appendix) shows
that both equations are equivalent. Notice that the solutions that are
analytic in $[0,1]$ in the transformed equation correspond to
solutions that are analytic in $[2,\infty]$ in the original equation,
and not to the physical modes. However, such solutions can be
generated from the physical ones by conformal transformations on the
two sphere\footnote{See \cite{heunsol} for an exhaustive list of Heun
  solutions and their relations.}, so both types appear for the same
values of the parameters. Notice that the both solutions have a
similar analytic structure, the only singularity is a branch cut
joining two of the singular points. Also the fact that the auxiliary
parameters $Q$ of both equations are the same for the particular cases
we are considering, allows an immediate identification of the complex
momentum numbers.

The transverse channel has different spectrum, whose first modes are
in table~\ref{tab:gluebVEC}. Although the glueballs associated to
vector fields have non-zero R-charge, and are usually not considered,
our computation shows that the lightest three-dimensional state and
hence, the longest correlation length, belongs to this
class.\footnote{This state is even lighter than the lightest QCD$_3$
  glueball listed in \cite{Brower:2000rp}.}
\TABLE[p]{
  \begin{tabular}{|c|r|} \hline\hline $n$ & $q_n^2$ \hspace*{3.5ex} \\
    \hline\hline \texttt{1} & $-5.1313$ \\ \hline \texttt{2} &
    $-22.4816$ \\ \hline \texttt{3} & $-51.2098$ \\ \hline \texttt{4}
    & $-91.4106$ \\ \hline \texttt{5} & $-143.0926$ \\ \hline
    \texttt{6} & $-206.2577$ \\ \hline \texttt{7} & $-280.9066$ \\
    \hline \texttt{8} & $-367.0395$ \\ \hline \texttt{9} & $-464.6566$
    \\ \hline \texttt{10} & $-573.7580$ \\ \hline
  \end{tabular}
  \caption{\label{tab:gluebVEC}The first ten glueballs of the
    transverse mode.}}

\subsection{Stress-energy tensor} \label{sec:metric} The stress-energy
tensor of the gauge theory encodes important dynamical and
thermodynamical properties of the plasma. Correlation functions of the
stress-energy tensor \begin{equation}G_{\mu\nu,\rho\sigma}
  (t-t',\bx-\by) =-i \,\theta(t-t') \,\vev{
    [T_{\mu\nu}(t,\bx),T_{\rho\sigma}(t',\by)] } ~,
\end{equation}are related to perturbations of the metric that leave the ${\rm
  S}^5$ factor invariant. Therefore, we want to introduce a small
fluctuation of the four-dimensional part of the metric $g_{\mu\nu}
=g_{\mu\nu}^0 +h_{\mu\nu}$.

In the gauge theory, the breaking of Lorentz symmetry to rotational
symmetry by temperature splits the Green's functions in transverse,
shear and sound channels, that in the zero frequency limit contain the
$J^{PC}=2^{++}$, $1^{++}$ and $0^{++}$ glueball spectrum. This is
reflected in the gravity dual, where the perturbations fall into three
different classes with decoupled field equations
\cite{Policastro:2002se, Policastro:2002tn, Kovtun:2005ev}. The
associated spin to each of these channels is also 2, 1 and 0, so we
will refer to them also as tensor, vector and scalar.

In the shear and sound channels there are also hydrodynamical modes
that describe the diffusion of conserved momentum and the propagation
of sound. We will not study the sound channel, but we will show that
complex momentum modes also capture the shear pole.

We will work with gauge-invariant variables, following
\cite{Kodama:2003jz}. There, the authors consider general metrics of
the form \begin{equation}\dd s^2=-\mathsf{F}(r) \,\dd t^2 +\frac{\dd
    r^2}{\mathsf{F}(r)} +r^2 \dd\sigma_n^2 ~, \end{equation}where
$\dd\sigma_n^2$ corresponds to a metric of a $n$-dimensional space of
constant sectional curvature $K=0,\pm 1$, and \be
\mathsf{F}(r)=K-\frac{2M}{r^{n-1}}-\lambda r^2 ~.  \end{equation}In
our case, $K=0$, $n=3$, $\lambda=-1$ and $M=r_{\rm H}^4/2$.

The Einstein equations are decomposed in tensor, vector and scalar
components relative to the three-dimensional metric. It is thus
possible to define three different gauge-invariant quantities to which
we can associate a Schr\"odinger-like equation of motion
\cite{Kodama:2003jz}. In the $z$ coordinate it reads \begin{equation}
  -f(z)\,\frac{\dd}{\dd z} \left( f(z)\,\frac{\dd\psi_I}{\dd z}
  \right) +V_I(z)\,\psi_I =\omega^2\,\psi_I ~,\quad I\equiv\{T,V,S\}
  ~, \end{equation} where for each perturbation we will have a
different potential.  Rewriting the Schr\"odinger equation by shifting
$(\omega,q) \mapsto r_{\rm H}(\omega,q)$, the potentials $V$ are given
by \begin{eqnarray}
  V_T(z) &=& \frac{f(z)}{4z^2} \,(15 +4\,q^2z^2 +9z^4 ) ~, \nn \\
  V_V(z) &=& \frac{f(z)}{4z^2} \,(\,3 +4\,q^2z^2 -27z^4) ~, \\
  V_S(z) &=& \frac{f(z)}{4z^2} \,\frac{1}{(1+6q^{-2}z^2)^2} \left( -1
    +4\,q^2z^2 +9z^4 +156z^6 -108\frac{z^2}{q^2}
    +540\frac{z^4}{q^4}+324\frac{z^8}{q^4} \right) ~, \nn
\end{eqnarray} for tensor, vector and scalar perturbations,
respectively.

By making the change of variable $x=1-z^2$, the equations for tensor
and vector perturbations lead to a Heun equation. For scalar
perturbations the situation is not so simple, and it requires a
separate analysis that we leave for future work, so in the following
we will be concerned only with tensor and vector perturbations.

\paragraph*{Tensor perturbations.} The characteristic exponents are
\[
\left(0;-i\frac{\omega}{4},i\frac{\omega}{4}\right) ~,\quad
\left(1;-\frac 3 4,\frac 5 4\right) ~,\quad
\left(2;-\frac{\omega}{4},\frac{\omega}{4}\right) ~,\quad
\left(\infty;\frac 3 4,\frac 3 4\right) ~.
\]
\begin{equation}\psi_T(x)=x^{-i \omega/4} (x-1)^{5/4}
  (x-2)^{-\omega/4} y(x) ~.
\end{equation}\begin{eqnarray}
  \alpha\beta &=& \left( \frac{\omega}{4}(1+i)-2\right)^2 ~,\quad Q =
  \frac{q^2}{4} +4-(2-i) \left((-1+3i)\frac{\omega}{4} +\frac{\omega^2}{8} 
  \right) ~, \nonumber \\
  \gamma &=& 1 -i\,\frac{\omega}{2} ~,\quad \delta=3 ~,\quad \epsilon
  =1-\frac{\omega}{2} ~.  \end{eqnarray}
\paragraph*{Vector perturbations.}  The characteristic exponents are
\[
\left(0;-i\frac{\omega}{4},i\frac{\omega}{4}\right) ~,\quad
\left(1;-\frac 1 4,\frac 3 4\right) ~,\quad
\left(2;-\frac{\omega}{4},\frac{\omega}{4}\right) ~,\quad
\left(\infty;-\frac 3 4,\frac 9 4\right) ~.
\]

\begin{equation}\psi_V(x)=x^{-i \omega/4} (x-1)^{3/4}
  (x-2)^{-\omega/4} y(x) ~.
\end{equation}
\begin{eqnarray}
  \alpha\beta &=& \frac{\omega}{4}(1+i)\left( \frac{\omega}{4}(1+i)-3 \right)
  ~,\quad Q =\frac{q^2}{4} -(1+5i)\frac{\omega}{4} -(2-i) \frac{\omega^2}{8} ~,
  \nonumber \\
  \gamma &=& 1 -i\,\frac{\omega}{2} ~,\quad \delta=2 ~,\quad \epsilon
  =1-\frac{\omega}{2} ~.  \end{eqnarray}
Notice that tensor fluctuations obey the same equations as a massless
scalar field, so the first modes of the spectrum are plotted in figure
\ref{fig:scalarqRw}.  The Heun equation we have for the vector
perturbations goes over to the one the authors in \cite{Nunez:2003eq}
found for the shear mode after the transformation described in the
appendix.

As we have commented above, vector fluctuations correspond to the
shear channel of the gauge theory. This channel is associated to the
momentum of the plasma, that as a conserved quantity is not absorbed
but diffused. In the hydrodynamical limit it is possible to find an
analytic expression for the diffusion pole \cite{Policastro:2002se}
\begin{equation}
  \omega= -i \,\frac{q^2}{4 \pi T} ~.
\end{equation}
We find good numerical agreement for this mode $q= (1+i)\sqrt{2
  \omega}$, as can be seen in fig.~\ref{fig:hydrometr}.

The results for the shear mode are shown in figure
\ref{fig:shearwthomega}. The real and imaginary parts of the five
lowest complex momentum are plotted as a function of the frequency.
Again, we find a similar behavior to scalar and vector modes. The
imaginary parts start out at a finite value at $\omega=0$, develop a
shoulder that is more pronounced for the higher modes and then fall
off rather fast until they enter a regime of slow decrease for large
frequencies. The real parts start out the $q^R =0$ for zero frequency.

\subsubsection{Glueball masses} \label{ssec:glueballsMET} We can find
the glueball spectrum of the effective three-dimensional theory by
taking the static limit $\omega=0$. Again we have to use the recursion
relation starting $n=1$ \erf{eq:r1} since $\alpha(\omega=0)=0$. The
results for the shear channel are compiled in
table~\ref{tab:gluebMET}. The glueball spectrum for neutral glueballs
has been computed using a similar supergravity approach in
\cite{Brower:2000rp}. The numbers we find differ actually somewhat
from the ones quoted in \cite{Brower:2000rp} for the $1^{++}$
glueballs. We attribute this to the different numerical methods that
have been used to obtain them.
\TABLE[p]{
  \begin{tabular}{|c|r|} \hline\hline $n$ & $q_n^2$ \hspace*{3.5ex} \\
    \hline\hline \texttt{1} & $-18.6758$ \\ \hline \texttt{2} &
    $-47.4951$ \\ \hline \texttt{3} & $-87.7228$ \\ \hline \texttt{4}
    & $-139.4167$ \\ \hline \texttt{5} & $-202.5882$ \\ \hline
    \texttt{6} & $-277.2408$ \\ \hline \texttt{7} & $-363.3762$ \\
    \hline \texttt{8} & $-460.9949$ \\ \hline \texttt{9} & $-570.0974$
    \\ \hline \texttt{10} & $-690.6838$ \\ \hline
  \end{tabular}
  \caption{\label{tab:gluebMET}The first ten glueballs of the shear
    mode.}}

\section{Conclusions and Outlook}  \label{sec:conclude}
We have established a relation between solutions to linearized field
equations with complex momenta in an AdS-black hole background and the
absorption lengths of a conformal gauge theory in a plasma phase. We
have explicitly studied some simple examples corresponding to scalar,
vector and metric fluctuations. Due to conformal symmetry, all
absorption lengths scale simply with $T^{-1}$. At zero frequency we
find agreement with previous computations of the effective
three-dimensional glueball spectrum \cite{Csaki:1998qr,
  deMelloKoch:1998qs, Brower:2000rp}. However, we prefer in this paper
to interpret our results as screening lengths for static fields. This
interpretation has also recently and independently been proposed in
ref. \cite{Bak:2007fk}.

Furthermore, we have computed the dependence of the absorption length
on the frequency. The results for the first modes are compiled in
figures \ref{fig:scalarqRw}, \ref{fig:longwthomega},
\ref{fig:transvwthomega} and \ref{fig:shearwthomega}. In all the
cases, the plasma is less absorptive for higher frequencies.  The
complex wave numbers also capture the hydrodynamical behavior for
R-charge and momentum diffusion. Our numerical results are in
agreement with the simple analytic continuation of the dispersion
relation for the hydrodynamic modes. This is shown in figures
\ref{fig:hydrovec} and \ref{fig:hydrometr}.

One of the interesting results of our study is that the longest
screening length (the lightest ``glueball'' mass in the dimensionally
reduced theory) corresponds to a state with non-vanishing R-charge.
Such a state does not belong to the spectrum of the QCD$_3$ theory,
i.e. the mass gap of the effective three dimensional theory is not the
one of QCD$_3$! Glueball masses play an important role in the
determination of the Debye screening length. Here one studies the
glueball exchange between open strings in the AdS black hole
background. As has been pointed out in \cite{Bak:2007fk} the mass gap
by itself is not important for the Debye screening, because only
specific operators can couple to the open string. Since these open
strings are R-charge neutral, the low mass states with non zero
R-charge do not couple to the string. However, the string
configuration one considers usually has its endpoints fixed on one
point on the ${\rm S}^5$ and it is also possible to consider strings
that end on different points on the ${\rm S}^5$. In such a situation
the light non-zero R-charge states might become relevant and could
modify the result for the screening length.

In this paper we have only studied the cases that can be reduced to
Heun equations and allow the application of the efficient continued
fraction approach to the calculation of the complex momentum
eigenvalues. It would certainly be interesting to extend the present
investigations to the cases that cannot be reduced to Heun equations.
In these cases one has to resort to elementary method of Frobenius
expansions and this slows down the numerical calculation considerably.
Nevertheless we think that this is an interesting problem especially
in view of the comparison to the glueball mass calculations.

Another rather intersting point is the question wether the absorption
length diverges in the limit of infinite frequency or wether it stays
finite. Unfortunately so far we know only about numerical methods to
evaluate the absorption lengths.

A related problem is the calculation of the the absorption lengths in
non-conformal holographic theories. Due to the presence of an
underlying scale the dependence on the frequency is likely to show a
more complicated pattern than the one we have found for the conformal
case in this paper. It will also be of high interest to compute
absorption lengths for the meson states that appear in theories with
D7-brane embeddings in the AdS black hole using the same methods that
have been employed in the study of meson quasinormal modes in
\cite{Hoyos:2006gb}. In \cite{Myers:2007we} it has recently been
emphasized that instabilities arise for near critical black hole
embeddings. Such instabilities show up as quasinormal modes with
positive imaginary part. As we have seen, similar instabilities can
also arise in the study of the absorption lengths. Since the
instabilities in the screening lengths arise at $\omega=0$ and for
real values of $q^2$ it might be much easier to search for these
instead of unstable quasinormal modes!

We hope to make progress on these questions in future research.

\acknowledgments The research of K.\,L. is supported by the Ministerio
de Ciencia y Tecnolog\'{\i}a through a Ram\'on y Cajal contract. The
research of S.\,M. is supported by an FPI 01/0728/2004 grant from
Comunidad de Madrid. I.\,A. , K.\,L. and S.\,M. are supported in part
by the Plan Nacional de Altas Energ\'{\i}as FPA-2006-05485 and EC
Commission under grant MRTN-CT-2004-005104. S.\,M. wants to thank the
Physics Department at Swansea University for very warm hospitality.
S.\,M. also wants to thank G. S\'anchez for her support. We would like
to thank G. Aarts, J. Barb\'on, P. Kumar, E. L\'opez, J. Mas and 
R. Schiappa for useful discussions.

\newpage
\appendix
%
\section{Effective potentials}\label{app:effectiveV}
In section \ref{ssec:locCMMs} we have presented the stability analysis
for a scalar field but it can be generalized for any field component
$\varphi(z)$ satisfying a decoupled linear second order differential
equation
\begin{equation}
  \varphi''(z) + A_1(z)\,\varphi'(z) + A_0(z)\,\varphi(z) +B(z)^2 \omega^2
  \varphi(z)=0 ~.
\end{equation}
Factorizing $\varphi(z) = \sigma(z) \,\phi(z)$ and normalizing the
$\phi''$ term
\begin{equation}
  \phi'' + \left( 2\frac{\sigma'}{\sigma}+A_1 \right)\phi' + \left( A_0
    +A_1\,\frac{\sigma'}{\sigma} +\frac{\sigma''}{\sigma} \right) \phi + B(z)^2
  \omega^2\,f=0 ~.
\end{equation}
We now change $B(z) \dd z =\dd z_*$ and divide by $B(z)^2$
\begin{equation}
  \partial_{z_*}^2 \phi +\omega^2 \phi + \frac{1}{B(z)}\left(
    2\frac{\sigma'}{\sigma}+A_1 + \frac{B'(z)}{B(z)} \right)\partial_{z_*} \phi + 
  \frac{1}{B(z)^2}\left( A_0 +A_1\,\frac{\sigma'}{\sigma}
    +\frac{\sigma''}{\sigma}\right) \phi =0~.
\end{equation}
This expression becomes a Schr\"odinger equation when $\sigma$
satisfies
\begin{equation}
  2\frac{\sigma'}{\sigma}+A_1 + \frac{B'(z)}{B(z)}=0~.
\end{equation}
Then, the same stability arguments can be applied with the proper
identification of the potential
\begin{equation}V_0(z)=-\frac{1}{B(z)^2}\left( A_0 +\frac{1}{4} \left(
      \left( \frac{B'}{B} \right)^2 -A_1^2 \right) -\frac{1}{2} \left(
      A_1' +\left( \frac{B'}{B} \right)' \right) \right) ~.
\end{equation}We will now apply this to the other equations under
consideration in this paper.
\begin{itemize}
\item transverse vector components
  \begin{equation}
    V_0 = \frac{f(z)}{4z^2} \left(  3 + 5 z^4 +4 q^2 z^2 \right)
  \end{equation}
\item longitudinal and temporal vector components
  \begin{equation}
    V_0 = -\frac{f(z)}{4z^2} \left( 1 + 7 z^4 -4 q^2 z^2\right)
  \end{equation}
\item gravitational vector perturbation (shear mode)
  \begin{equation}
    V_0 = \frac{f(z)}{4z^2} \left( 3 - 27 z^4 +4 q^2 z^2\right)
  \end{equation}
\end{itemize}
Due to the underlying analyticity of the solution of the corresponding
Heun equation all fields, $\bA_T, V_L, V_0, \Phi_V$, fulfill the
boundary conditions leading to \erf{eq:stability.im}
\erf{eq:stability.re}.  The effective potential is positive in the
case of the transverse Vector fields.  For the longitudinal and
temporal vector field components it is negative and therefore the
stability argument presented in section \ref{ssec:locCMMs} does not
apply. We note that the asymptotic behavior at the boundary is the
same as that of a scalar field saturating the Breitenlohner-Freedman
bound. We take this as an indication for stability, in the original
analysis in AdS a positive energy condition is satisfied even for
fields with negative potential \cite{Breitenlohner:1982bm}. The
asymptotic behavior of the fields is restricted by the condition of
having a well-defined conserved energy. In turn, the positive
contribution of the kinetic energy always overcomes the negative
contribution from the potential.
 A formal analysis \cite{Ishibashi:2003ap} can be applied that shows the stability of vector perturbations. The 'Hamiltonian' operator $H=-\partial_{z_*}^2+V_0$ must be positive definite over the set of normalizable solutions
\begin{equation}\label{eq:posenergy}
\int_{z_*^b}^\infty d z_* \chi^* H \chi >0 \ .
\end{equation}
We can rewrite (\ref{eq:posenergy}) as 
\begin{equation}
-\left[\chi^* D_\rho \chi \right]_{z_*^b}^\infty +\int_{z_*^b}^\infty d z_*\left(|D_\rho \chi|^2+V_\rho |\chi|^2\right)>0\ ,
\end{equation}
where we have introduced an auxiliary function $\rho$, such that $D_\rho=\partial_{z_*}+\rho$ and
\begin{equation}
V_\rho=V_0+\partial_{z_*} \rho -\rho^2\ .
\end{equation}
A convenient election that makes $V_\rho \geq 0$ for vector fluctuations is $\rho=-f(z)/2z$. We can easily see that there is no contribution from the boundary term at the horizon, since $\chi(\infty)\to\, {\rm const.}$ and $\rho(\infty)\to 0$. Therefore, we are left with the condition
\begin{equation}\label{eq:zerobound}
 \lim_{z\to 0} \chi^* \left(\partial_z -{1\over 2z}\right) \chi =0\ .
\end{equation}
Close to the boundary, $V_0\simeq -1/4 z^2$, so the solution is a combination of Bessel functions  $\chi \sim a \sqrt{z} J_0(\omega z) + b \sqrt{z} Y_0(\omega z)$. Then, the condition (\ref{eq:zerobound}) satisfied when $b=0$, that is equivalent to choose the normalizable
 solution at the boundary.

The effective potential of the shear mode is also interesting. It is
negative close to the horizon. Equation \erf{eq:stability.re} shows
that this is a necessary requirement for existence of the hydrodynamic
shear mode with $\Re(q^2)=0$. We would expect that if the potential is
deep enough, instabilities will appear.  This is in agreement with
other analysis that exhibit a negative well in the interior. Purely
imaginary frequencies have been found in the study of electromagnetic
and gravitational perturbations in global AdS
\cite{Cardoso:2001bb,Berti:2003ud}. In the extremal limit, the
frequencies seem to reach the real axis at $\omega=0$, and the
geometry was conjectured to be marginally unstable. Recent works also
suggest that instabilities of D7 probe branes in AdS appear when a
quasinormal mode cross the real axis at $\omega=0$
\cite{Hoyos:2006gb,Myers:2007we}.

\section{Changing parameters in a Heun equation} \label{app:Heuneq} In
this appendix we show how to map a given Heun equation with given
parameters into another Heun equation for a different function with a
different set of parameters. This will allow in some cases to avoid
the problem with ``fake" modes.

Let us start with a Heun equation for $y(x)$ \begin{equation}y''
  +\left( \frac{\gamma}{x} +\frac{\delta}{x-1} +\frac{\epsilon}{x-2}
  \right) y' +\frac{\alpha\beta x -Q}{x(x-1)(x-2)} ~y(x) =0 ~,
\end{equation}where the parameters are subject to the condition
$\alpha+\beta+1=\gamma+\delta+\epsilon$. The characteristic exponents
at the AdS boundary $(x=1)$ are in general
\[
\{1;0,1-\delta\} ~,
\]
for $y(x)$. In the cases where $\delta \in \{0 \,\cup \,\bbZ^-\}$, the
first solution is logarithmic, but the logarithm might accidentally
vanish at the ``false frequencies" \begin{equation}y(x) =B[ 1+(1-x)
  +\ldots +h (1-x)^{(1-\delta)}\log(1-x) +\ldots] +A[
  (1-x)^{(1-\delta)} +\ldots] ~, \end{equation}where $h=0$. According
to \cite{Nunez:2003eq}, the Green's function is proportional to the
ratio $A/B$, and the false frequencies are the ones for which
accidentally $h=0$.

Now we would like to change the function $y(x)$ such that we find a
related Heun equation with different parameters. Let us define \be
y(x) =(1-x)^\varrho \,Y(x) ~.  \end{equation}This allows us to find a
Heun equation for $Y(x)$, \textit{provided} $\varrho=1-\delta$, and
where the new set is \begin{eqnarray}
  \widetilde{\alpha} &=& \alpha+\varrho =\alpha+1-\delta ~, \nn \\
  \widetilde{\beta} &=& \beta+\varrho =\beta+1-\delta ~, \nn \\
  \widetilde{Q} &=& Q +2\gamma\varrho =Q +2\gamma(1-\delta) ~, \\
  \widetilde{\delta} &=& \delta+2\varrho =2 -\delta ~, \nn \\
  \widetilde{\gamma} &=& \gamma ~,\quad \widetilde{\epsilon} =\epsilon
  ~. \nn \end{eqnarray} The interesting thing about this shift is that
we can find a positive $\widetilde{\delta} \in\bbZ^+$ when in the
original Heun equation we encounter fake frequencies. This always
eliminates the false frequencies since the second solution is never
analytic but goes like $(1-x)^{(\delta-1)}$, which is a negative (or
zero) exponent for $\widetilde{\delta} \in \bbZ^+$. Now, the Frobenius
solution is
\begin{equation}
  Y(x) =B [ (1-x)^{-\varrho} +\ldots
  +h(1-x)^{(1-\delta-\varrho)}\log(1-x) +\ldots] +A[
  (1-x)^{(1-\delta-\varrho)} +\ldots] ~.  
\end{equation}
The recursion algorithm of Leaver \cite{Leaver:1985ax} and its
adaption to the Heun equation by Starinets \cite{Starinets:2002br}
computes when the solution of the Heun equation is analytic at $x=1$.
Now the solution that goes with the coefficient $B$ is never analytic,
and therefore the false frequencies do not appear.

\newpage


\newpage
\bfig[!htbp]
\centering
\includegraphics[scale=0.92]{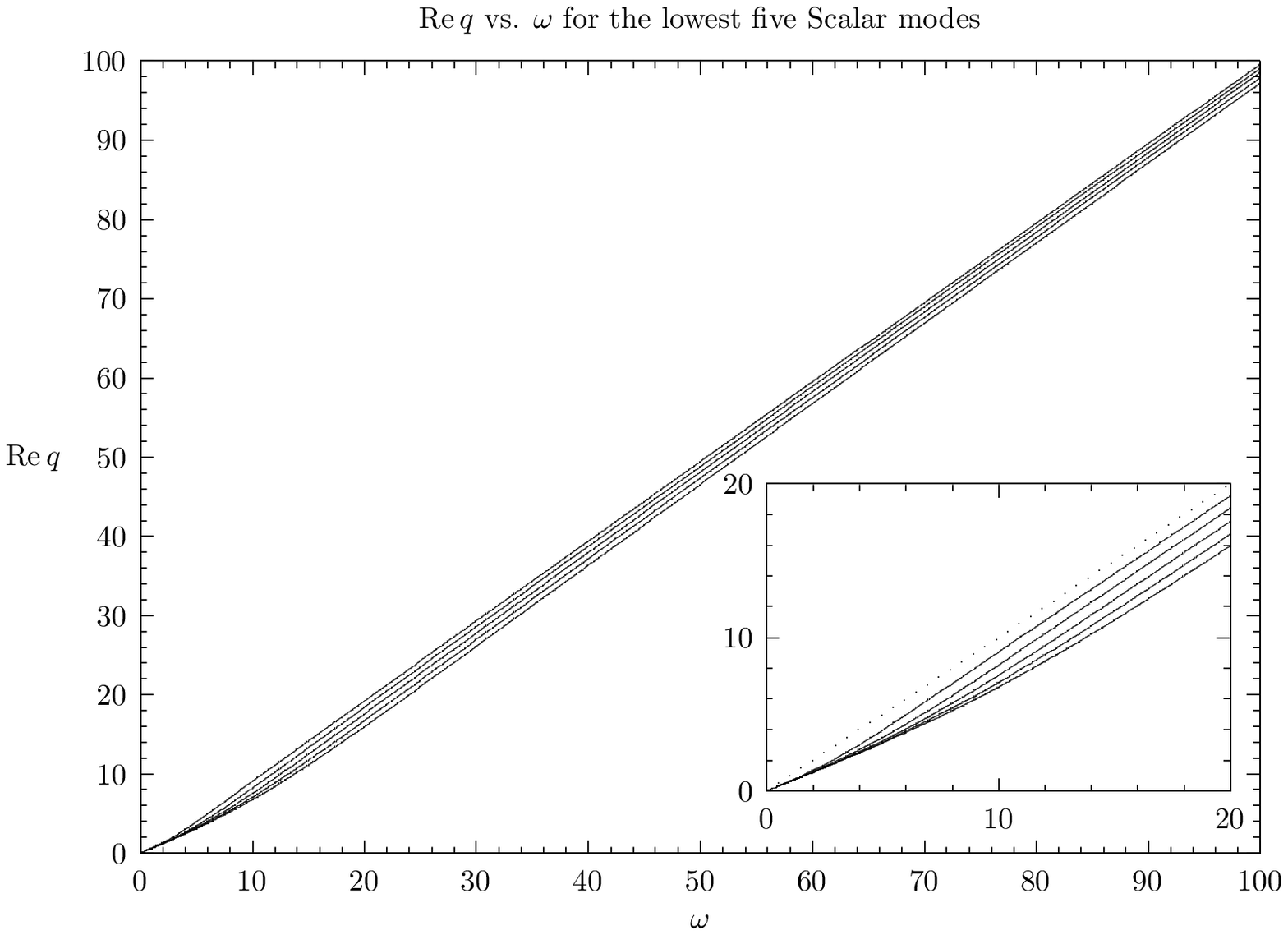} 
\centering
\includegraphics[scale=0.92]{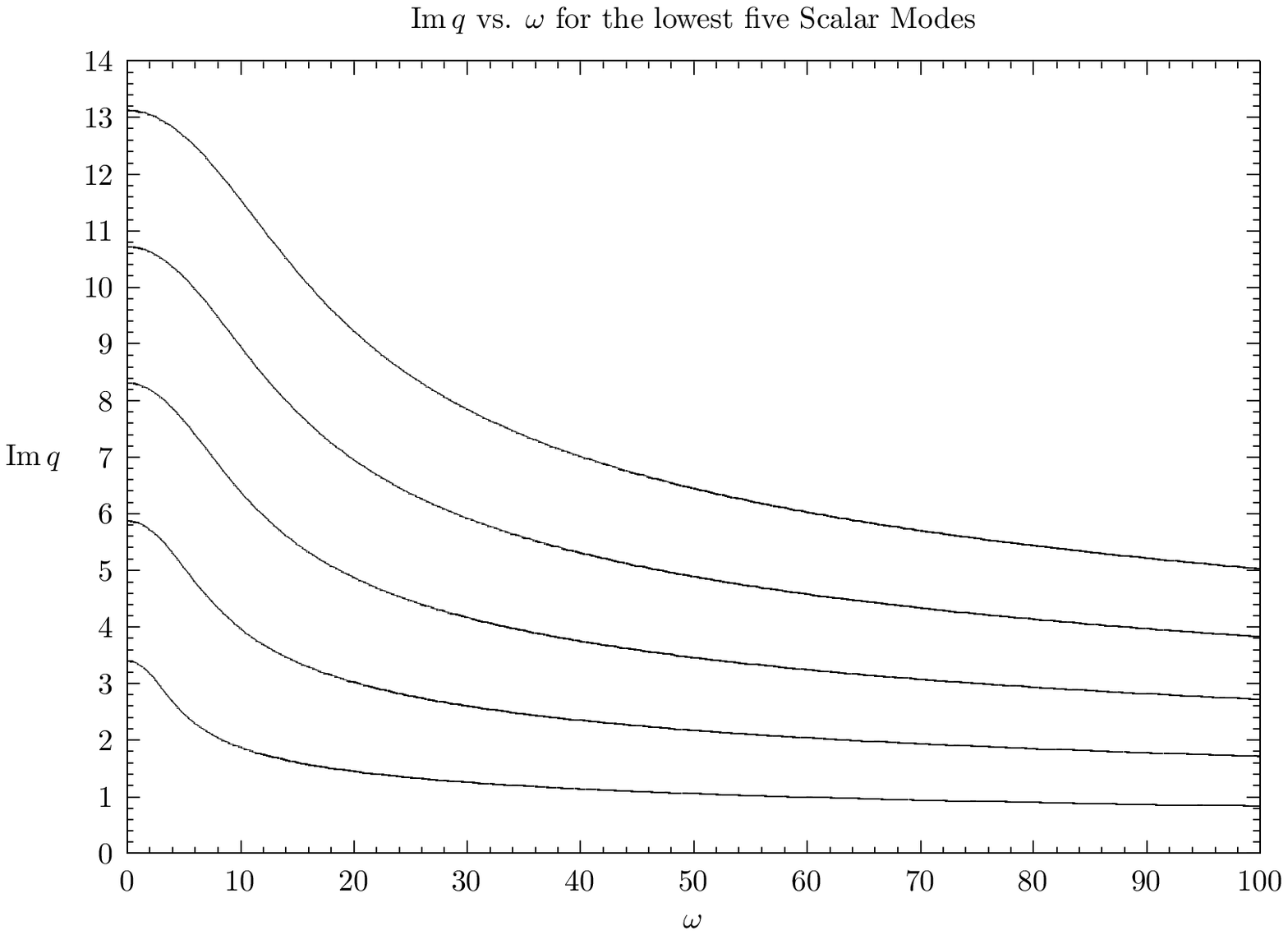}
\caption{\label{fig:scalarqRw}Real and imaginary parts of the lowest five complex momentum eigenvalues
versus the frequency. In the lower-right corner of the first figure we have
zoomed in to show the separation between the five modes.}
\efig

\newpage
\bfig[!htbp]
\centering
\includegraphics[scale=0.86]{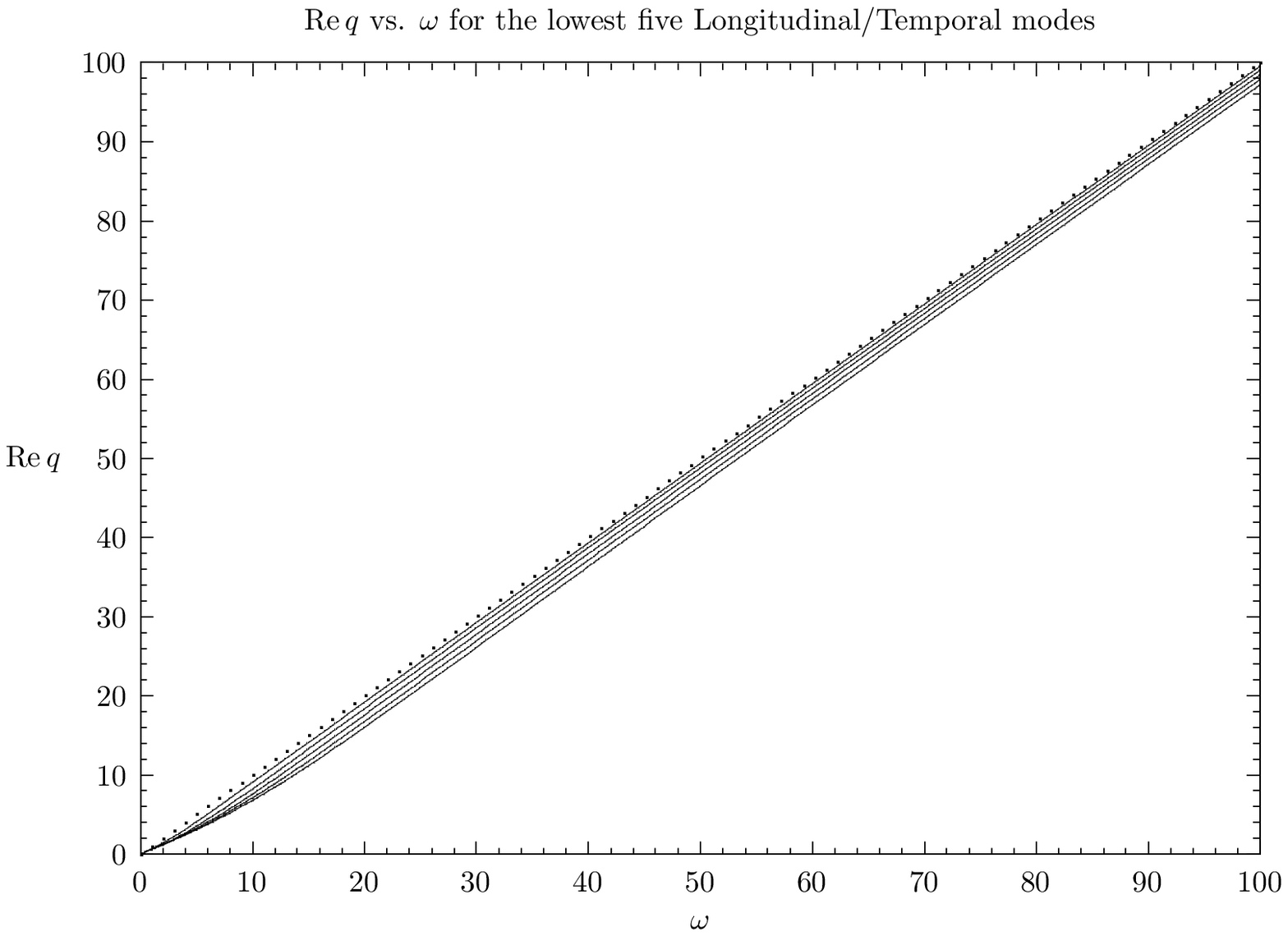}
\includegraphics[scale=0.86]{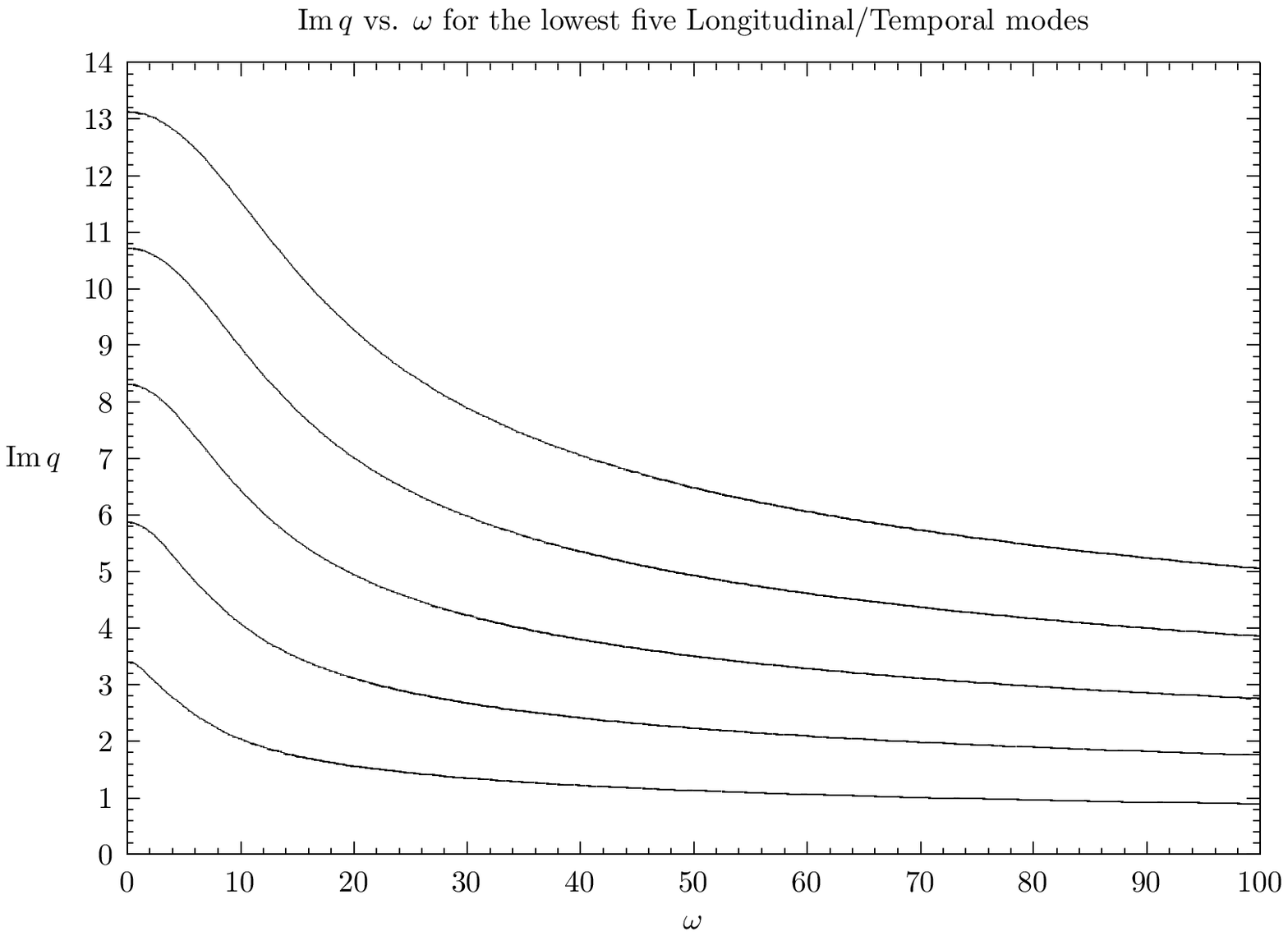}
\caption{\label{fig:longwthomega}Longitudinal perturbations of the 
vector field on AdS. At $\omega=0$ the values coincide with the ones of 
the scalar field perturbations. For $\omega>0$ the shape is however
different.}
\efig

\newpage
 \bfig[!htbp]
\centering
 \includegraphics[scale=0.86]{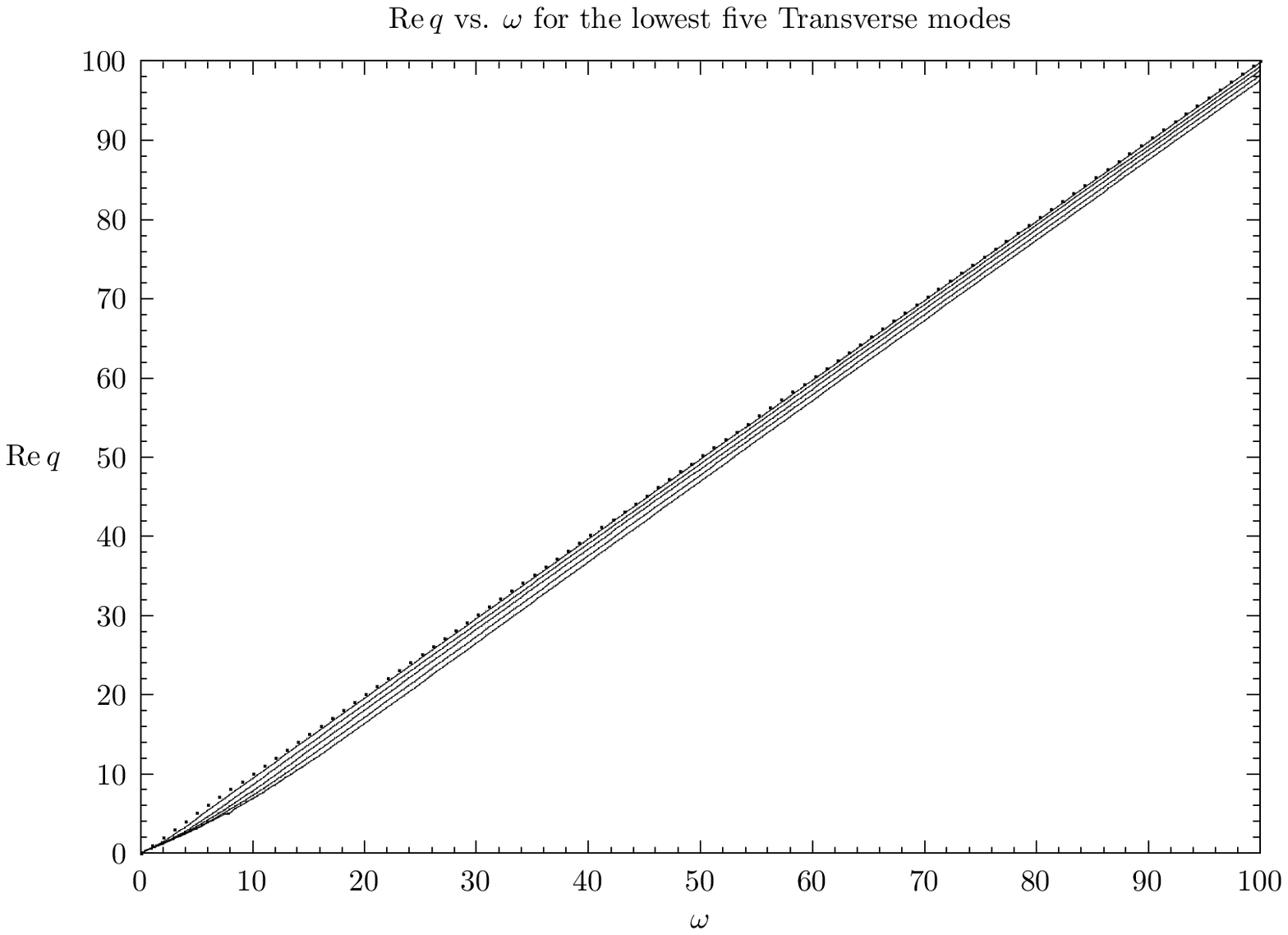}
 \includegraphics[scale=0.86]{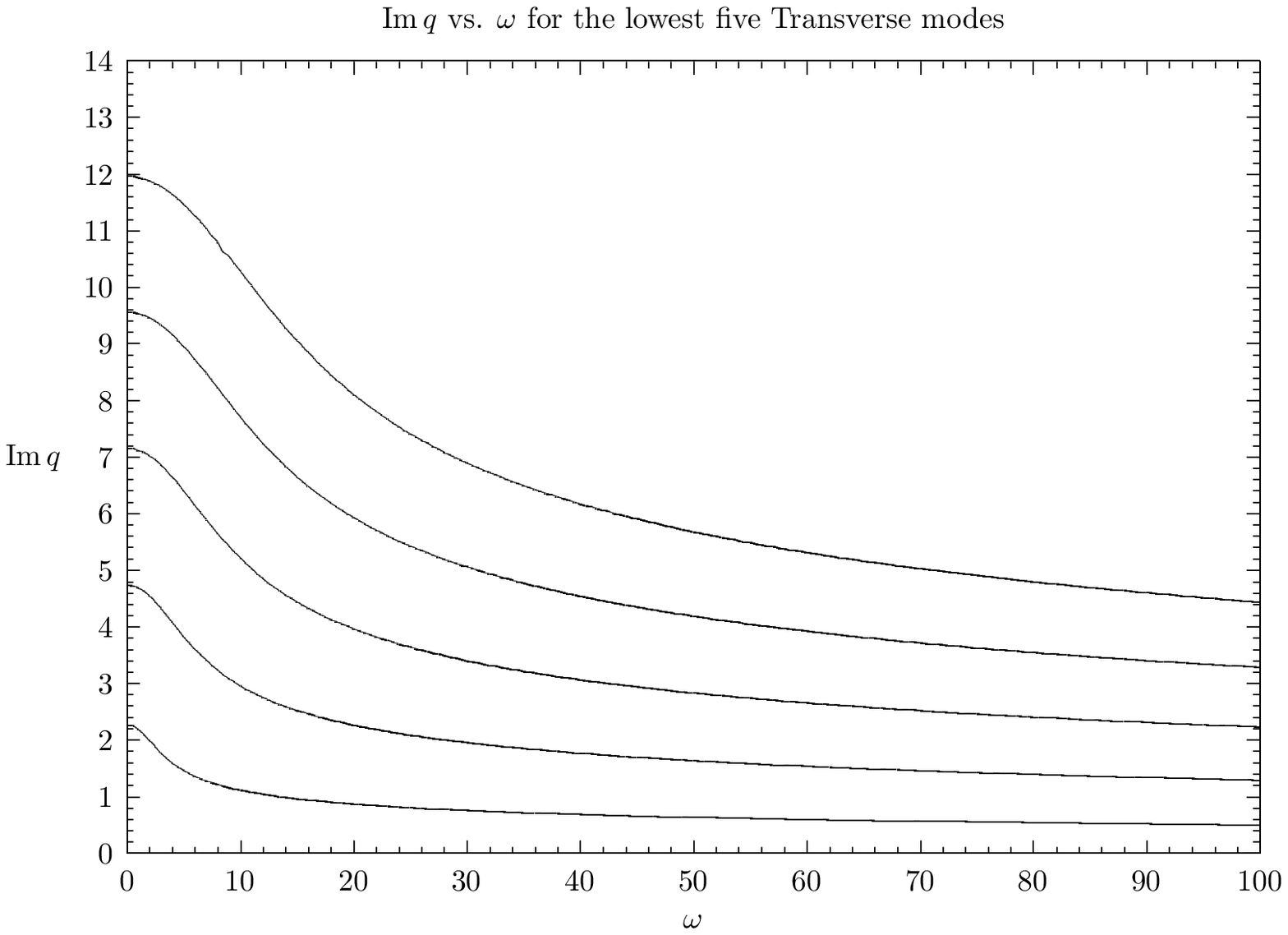}
 \caption{\label{fig:transvwthomega} The complex momentum 
eigenvalues of the transverse vector field components. Note that
the lowest mode gives the longest absorbtion length. The plasma is
most transparent to transverse vector perturbations.}
\efig

\newpage
\bfig[!htbp]
\centering
\includegraphics[scale=0.86]{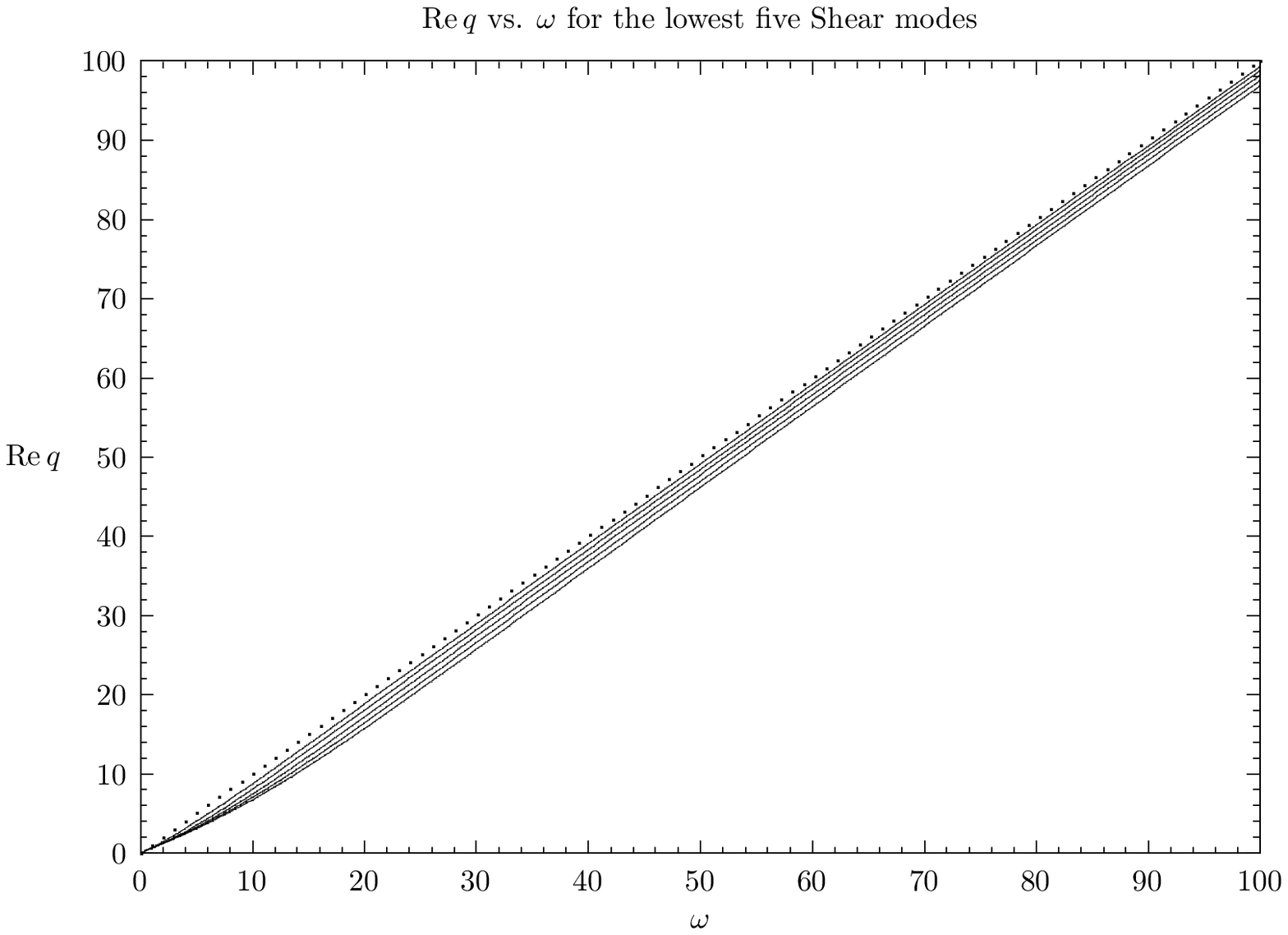}
\includegraphics[scale=0.86]{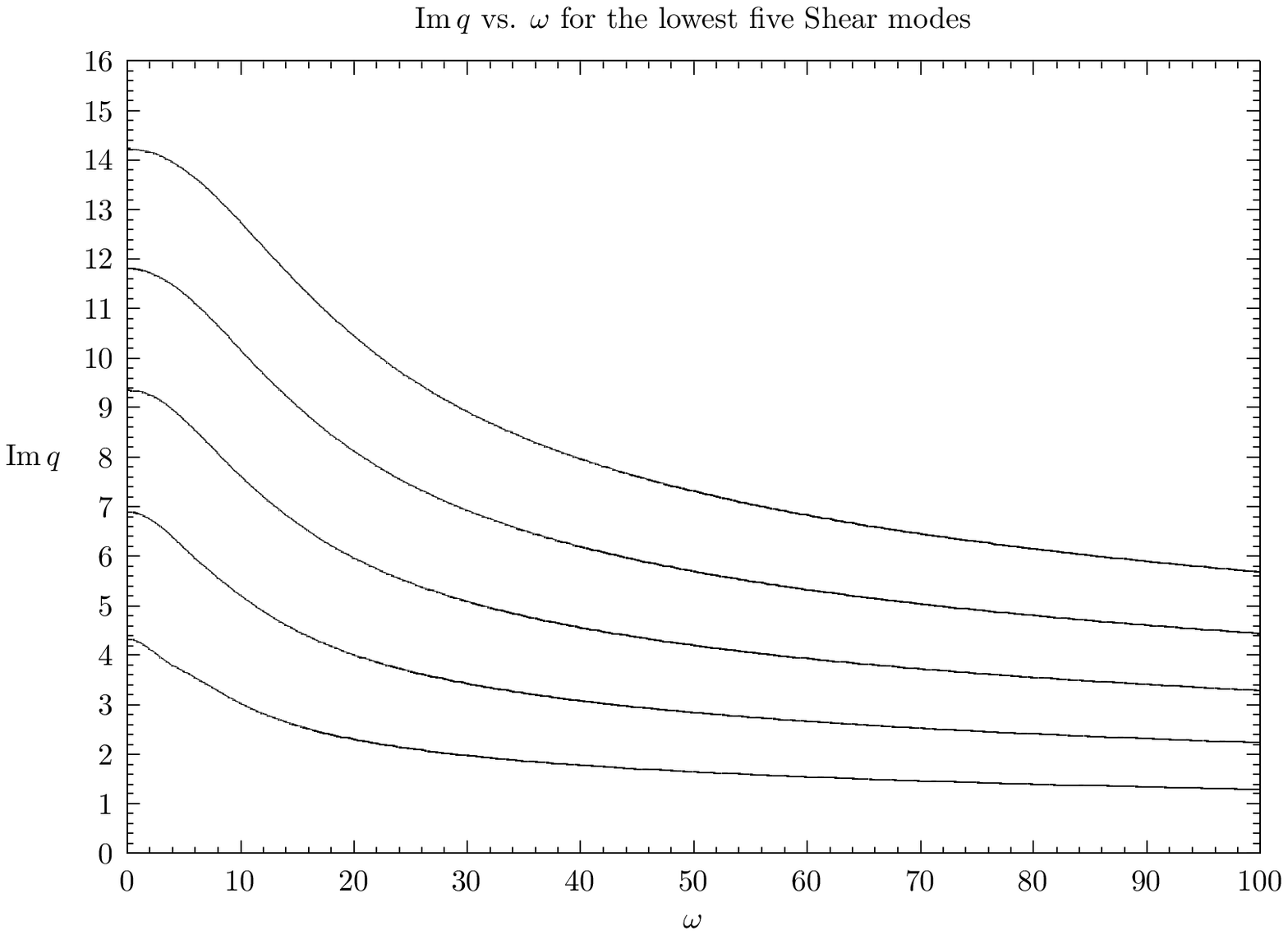}
\caption{\label{fig:shearwthomega} The lowest five complex momentum 
eigenvalues above the diffuse mode for the shear channel perturbations.}
\efig

\newpage
\bfig[!htbp]
\centering
\includegraphics[scale=0.86]{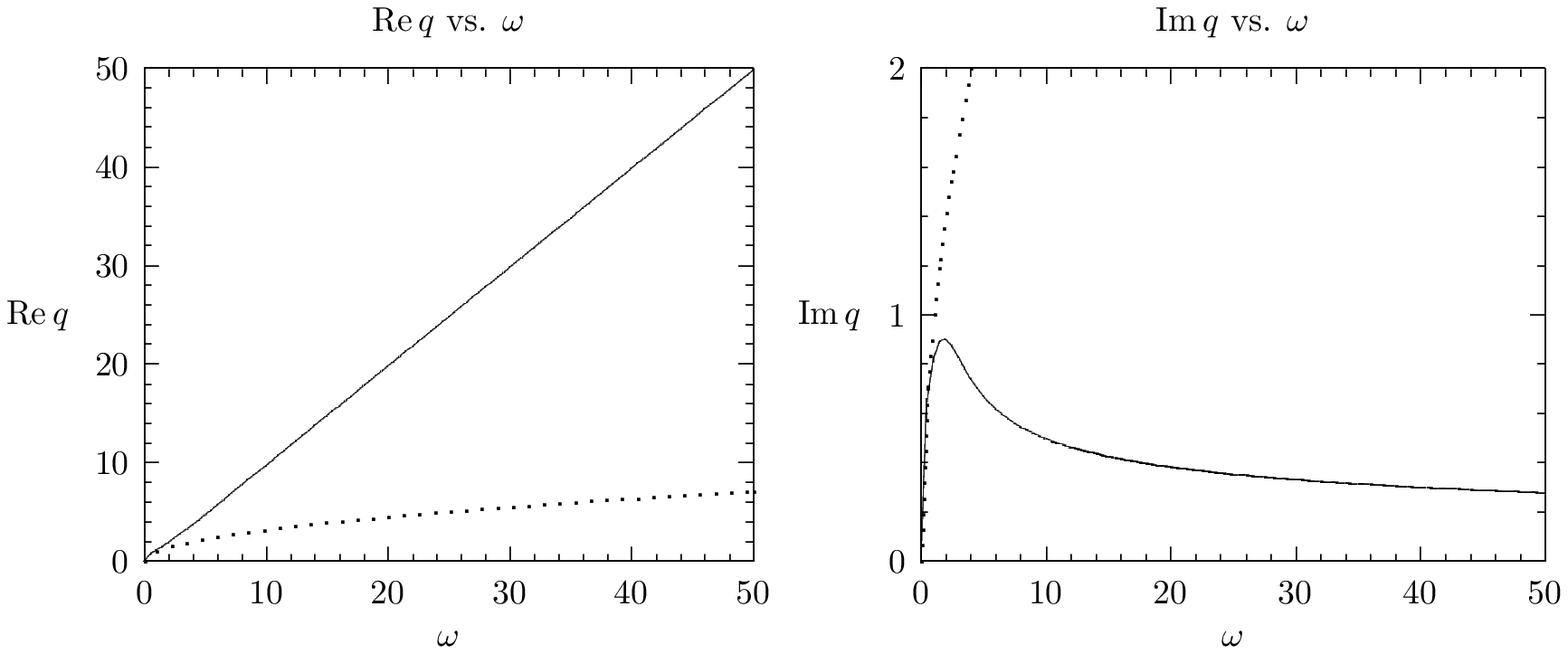}
\caption{\label{fig:hydrovec} Diffusion mode. The solid line represents the
numerical solution while the dotted line is the analytic formula from
hydrodynamical analysis.}
\efig

\bfig[!htbp]
\centering
\includegraphics[scale=0.86]{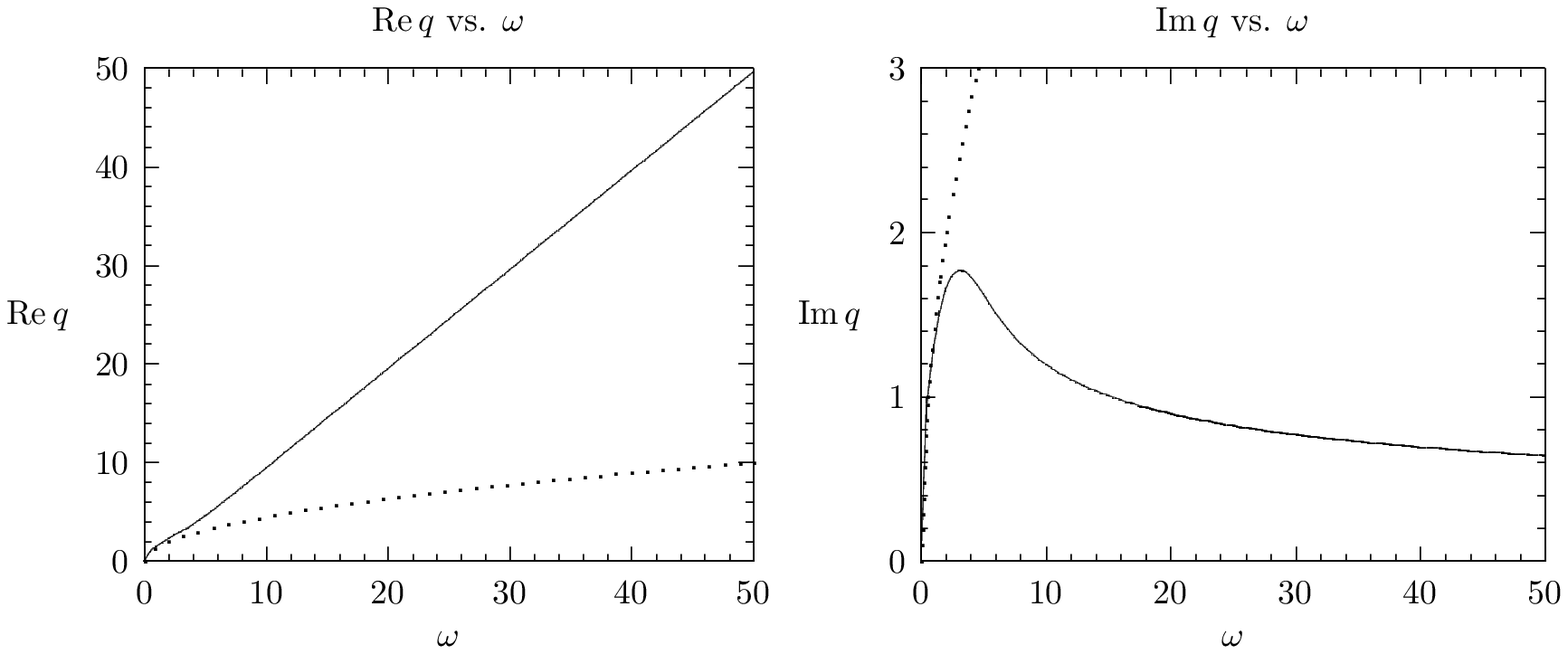}
\caption{\label{fig:hydrometr} Shear mode. The solid line represents the
numerical solution while the dotted line is the analytic formula from
hydrodynamical analysis.}
\efig

\end{document}